\documentclass[a4paper,english,english,aps,pre,singlecolumn,eqsecnum]{revtex4}
\usepackage[T1]{fontenc}
\usepackage[latin9]{inputenc}
\usepackage{babel}
\usepackage{array}
\usepackage{verbatim}
\usepackage{multirow}
\usepackage{amsmath}
\usepackage{amssymb}
\usepackage{graphicx}
\usepackage{esint}
\usepackage[unicode=true,
 bookmarks=false,
 breaklinks=false,pdfborder={0 0 1},
 colorlinks=false]
 {hyperref}
\hypersetup{
 colorlinks,citecolor=cyan,linkcolor=blue,urlcolor=cyan}
\usepackage{breakurl}

\makeatletter

\providecommand{\tabularnewline}{\\}

\@ifundefined{textcolor}{}
{%
 \definecolor{BLACK}{gray}{0}
 \definecolor{WHITE}{gray}{1}
 \definecolor{RED}{rgb}{1,0,0}
 \definecolor{GREEN}{rgb}{0,1,0}
 \definecolor{BLUE}{rgb}{0,0,1}
 \definecolor{CYAN}{cmyk}{1,0,0,0}
 \definecolor{MAGENTA}{cmyk}{0,1,0,0}
 \definecolor{YELLOW}{cmyk}{0,0,1,0}
}

\newsavebox{\@brx}
\newcommand{\llangle}[1][]{\savebox{\@brx}{\(\m@th{#1\langle}\)}%
  \mathopen{\copy\@brx\kern-0.7\wd\@brx\usebox{\@brx}}}
\newcommand{\rrangle}[1][]{\savebox{\@brx}{\(\m@th{#1\rangle}\)}%
  \mathclose{\copy\@brx\kern-0.7\wd\@brx\usebox{\@brx}}}

\makeatother

\begin{document}

\title{Equilibration in one-dimensional quantum hydrodynamic systems}

\author{Spyros Sotiriadis}

\affiliation{Institut de Math\'{e}matiques de Marseille, (I2M) Aix Marseille Universit\'{e},
CNRS, Centrale Marseille, UMR 7373, 39, rue F. Joliot Curie, 13453,
Marseille, France} 
\affiliation{University of Roma Tre, Department of Mathematics and Physics, L.go
S. L. Murialdo 1, 00146 Roma, Italy}
\maketitle

\begin{quotation}
\begin{center}
\emph{Dedicated to John Cardy on the occasion of his 70th birthday} 
\par\end{center}
\end{quotation}

\textbf{Abstract}
We study quench dynamics and equilibration in one-dimensional quantum
hydrodynamics, which provides effective descriptions of the density
and velocity fields in gapless quantum gases. We show that the information
content of the large time steady state is inherently connected to
the presence of ballistically moving localised excitations. When such
excitations are present, the system retains memory of initial correlations
up to infinite times, thus evading decoherence. We demonstrate this
connection in the context of the Luttinger model, the simplest quantum
hydrodynamic model, and in the quantum KdV equation. In the standard
Luttinger model, memory of all initial correlations is preserved throughout
the time evolution up to infinitely large times, as a result of the 
purely ballistic dynamics. However nonlinear dispersion or 
interactions, when separately present, lead to spreading
and delocalisation that suppress the above effect by eliminating the 
memory of non-Gaussian correlations. 
We show that, for any initial state that satisfies sufficient
clustering of correlations, the steady state is Gaussian in terms
of the bosonised or fermionised fields {in the 
dispersive or interacting case} respectively. On the other
hand, when dispersion and interaction are simultaneously present, 
a semiclassical approximation suggests that localisation is restored
as the two effects compensate each other and solitary waves are formed.
Solitary waves, or simply solitons, are experimentally observed in
quantum gases and theoretically predicted based on semiclassical approaches,
but the question of their stability at the quantum level remains to
a large extent an open problem. We give a general overview on the
subject and discuss the relevance of our findings to general out of 
equilibrium problems.

\tableofcontents

\section{Introduction\label{sec:intro}}

A decade ago Calabrese and Cardy \cite{CC_QQ} introduced the concept
of \emph{quantum quench} as a generic physical setup for the study
of universal features in out-of-equilibrium quantum statistical physics.
Together with a profound conjecture for the large time steady state
in integrable models, the \emph{Generalised Gibbs Ensemble} (GGE)
of Rigol et al. \cite{Rigol_GGE} proposed in the same year, these
two concepts have been highly inspiring, especially as a means to
understand the mechanism of equilibration in closed quantum systems
\cite{Eisert,s-jstat1,s-jstat2,s-jstat3,Bernard-Doyon,QA2,s-jstat6,quasi-local2,s-jstat7}.
Interestingly, the celebrated first experimental observation of a
GGE that was only recently achieved \cite{exp:GGE} brings back the
original problem considered in \cite{CC_QQ}, that of the quantum dynamics
in a one-dimensional (1D) gapless system.

Perhaps the most intriguing problem regarding equilibration in quantum
systems is the mechanism of memory loss: Let us consider an extended
quantum system prepared in a general initial state and let it evolve
under unitary dynamics over a long time. Would it relax locally to
a statistical ensemble that retains comparatively little memory of
the initial state? What are the physical conditions and what the physical
mechanism of this memory loss? What presicely is the information content
of the steady state ensemble for a general initial state?

Integrable models are, in this respect, an ideal testing ground of
equilibration: on the one hand, they are characterised by an extensive
number of conservation laws that prevent relaxation to a thermal ensemble;
on the other hand, their exact solvability allows 
the \emph{ab initio} study of their dynamics, thus offering us
the opportunity to identify the physical principles behind memory
loss and memory preservation in out-of-equilibrium quantum systems.
What the GGE conjecture essentially postulates is that, while such
integrable systems do not thermalise, they still equilibrate locally
but to an ensemble that contains information about their extra conservation
laws. Even though it is still unclear which conserved quantities should
be included in the general case, in its currently most complete form
\cite{GGEfail1,GGEfail2,completeGGE} the GGE is defined through the
information of the distribution densities of distinct types of initial excitations,
which amounts to including both local and the recently discovered
quasi-local conserved charges \cite{quasi-local,quasi-local2}. Despite
the fact that the number of such charges in a thermodynamically large
system is infinite, this is still an enormously economic ensemble:
While a complete description of local observables in a general initial
state would require knowledge of each multi-point correlation function separately,
or equivalently, of the amplitudes of clusters of excitations of any order 
\cite{cluster}, the above ensemble contains information only about the 
distribution densities of excitations. Therefore relaxation to the GGE is
associated with an enormous memory loss, from an initial information
content that scales exponentially with the system size to just linear
scaling in the final state. This memory loss effect is expected to
be a consequence of dephasing due to the destructive interference
of quasiparticles originating from very distant initially points.
From this perspective gapless systems are somewhat special: the absence of
an energy gap means that there is no stationary phase point, which
is the most common cause of dephasing.

While the GGE conjecture has been verified in a large number of quantum
quenches, the general conditions for its validity, the set of conserved
charges it must generally contain and more importantly a deeper understanding
of its success, are still missing. One approach to prove it from first
principles of quantum many-body physics relates it to two precise
and broadly applicable physical conditions, one for the initial state
and one for the dynamics \cite{Cramer-Eisert,Cramer-Eisert_b,SC_GGE-val,FE_cluster-dec,S15,doyon,Eisert16}.
For the initial state the key factor is the \emph{clustering of correlations}
of local fields, a rather general property of quench initial states
as a consequence of the locality of the pre-quench Hamiltonian. The
role of initial clustering in equilibration was first demonstrated
in \cite{Cramer-Eisert_b} in the context of non-interacting lattice
models, where it was shown that any initial state that satisfies the
cluster property relaxes locally to a Gaussian GGE. 
The loss of memory is clearly reflected in the fact that a general 
non-Gaussian initial state relaxes to a Gaussian one that is completely determined by 
the knowledge of its two point correlation function. Apart from initial clustering, 
the proof relied also on the presence of a \emph{maximum velocity of propagation},
as rigorously expressed by the Lieb-Robinson bound of lattice systems
with finite-range couplings \cite{LR}. This approach has been recently
extended to the far more general case of genuinely interacting lattice systems
with finite-range couplings, including non-integrable ones \cite{doyon}.
In the context of continuous field theories, the importance of the
cluster property was rediscovered in \cite{SC_GGE-val}. The existence
of a maximum group velocity of excitations is by default a
property of relativistic quantum field theories, but it turned out
that it is not a necessary condition for the validity of the GGE, since the same
conclusion was reached also in a non-relativistic bosonic system.
On the other hand, in \cite{S15} it was shown that when the evolution
follows a free massless relativistic field theory, despite the inherent
light-cone dynamics, the equilibrium state is not described by a Gaussian
GGE, as memory of all initial correlation functions is preserved up
to infinite time. Back to the lattice, a new perspective on the equilibration
mechanism was introduced in \cite{Eisert16} by unveiling a link to
transport properties. It was discovered that a sufficient condition
on the dynamics is that the evolution Hamiltonian exhibits \emph{delocalised
transport}, i.e. spreading of initially localised wavepackets such that they 
decay with time uniformly in the whole space. Even though all above results, with the exception of \cite{doyon},
were demonstrated under the assumption of essentially non-interacting
dynamics, the conclusions of \cite{doyon} suggest that they naturally
extend to the interacting case.

The condition of delocalised transport can be easily understood intuitively:
if the time evolution follows a Hamiltonian characterised by localisation,
then the initial state and its correlations practically ``freeze''
and do not evolve significantly \cite{theor:MBL-dynamics}. In fact
this effect was exploited for the experimental observation of many
body localisation induced by disorder \cite{exp:MBL}. Though seemingly
contraddictory, a not so different type of localisation occurs also
in systems that exhibit purely ballistic transport: those for which
the Heisenberg equations of motion admit ballistically moving localised
solutions. Gapless relativistically invariant systems described by
the standard Luttinger model with linear dispersion 
are the simplest examples of such type of transport.
Due to the purely ballistic nature of transport, entangled clusters
of left or right-moving particle excitations of the initial state
are parallel transported throughout the time evolution, carrying their
initial correlations intact up to infinite times \cite{S15}.

The demonstration of the above memory effect in the Luttinger model
relies on its gaplessness and relativistic invariance. While this
model has been argued to provide a good description of the dynamics
in relevant cold atom experiments \cite{exp:lightcone1,exp:lightcone2},
the ultimate fate of this memory effect depends on whether deviations
of physical systems from the Luttinger model undermine the stability
of such ballistic localised excitations. The most important deviations
in quantum gases are the nonlinear dispersion and the presence of
interactions that are not captured by the standard Luttinger model
\cite{nonlinear_LL1,nonlinear_LL}. We show that each of these features,
when considered separately, breaks the purely ballistic character
of the dynamics, leading to spreading of wavepackets and delocalisation.
Due to this effect, in analogy with \cite{Eisert16} we show that
in both of these cases equilibration is characterised by memory loss
and the steady state is a Gaussian GGE in terms of a suitable choice
of fields. On the other hand, in the presence of both dispersion and
interactions their combined effect opens up another possibility: Let
us recall that in classical integrable models\emph{ solitary waves,}
or simply \emph{solitons}, emerge as the result of mutual cancellation
between dispersion and interaction and notice that solitary waves
are characterised by precisely the features we have identified as
leading to memory preservation, i.e. ballistic transport and localisation.
As observed in \cite{GP-solitons,classical_soliton-quench}, we 
find that a semiclassical approach would suggest that
such solitons emerge and restore the ballistic localised character
of the dynamics, raising the question whether this type of behaviour persists when
quantum fluctuations are taken into account.

This brings to the forefront the emerging field of 1D quantum hydrodynamics
\cite{q_hydrodyn1,QH0,QH1,QH2,QH3,QH4,GP-solitons_long,q_hydrodyn2,Bernard-Doyon2} and,
as a prototype model, the \emph{quantum Korteweg-de Vries} (KdV) equation
\cite{qKdV-LL,qKdV1,qKdV2,GP-solitons,GP-solitons_long}, which is expected to capture
the leading deviations from Luttinger model that are relevant
to the dynamics of 1D Bose gases \cite{qKdV-LL,GP-solitons}. Experiments
in 1D cold atomic gases have achieved the observation of dispersionless
bright solitons \cite{exp:bright_soliton} and the formation and propagation
of soliton trains \cite{exp:soliton_train} in the case of attractive
interaction, as well as dark solitons that eventually disperse in
the case of repulsive interactions \cite{exp:dark_soliton}. The existence
of solitons in such quantum gases has been theoretically predicted
by means of the semiclassical approach, using the Gross-Pitaevski
or Non-Linear Schrödinger Equation (NLSE) \cite{theor:solitons_NLS},
the KdV \cite{KdV_solution,ZK} and the Benjamin-Ono equation \cite{B-O},
all of which admit classical soliton solutions. Stable soliton solutions
have been argued to persist at the quantum level in the quantum Benjamin-Ono
equation \cite{q_hydrodyn1}, which is the hydrodynamic description
of the Calogero-Sutherland model, confirming in this way earlier predictions
\cite{Jevicki,soliton_calogero,soliton_calogero2}. These findings
are evidence of the existence of solitary waves in quantum hydrodynamic
systems, whose presence in different models remains however an open
problem. More specifically, in the repulsive Lieb-Liniger model, whose low-energy
effective hydrodynamic description is given by the quantum KdV equation,
the semiclassical approach in combination with Bethe-Ansatz analysis
suggests that the solitons become unstable and disperse at large times,
even though with a weaker dispersion law than the phononic excitations
\cite{qKdV-LL}. In this model it has been conjectured that static
dark solitons can be formed out of Lieb's type-II excitations \cite{dark-solitons-Lieb2,classical-quantum_NLSE}
and numerical results seem to confirm this conjecture, although they
are not yet conclusive about their dynamical stability \cite{ds_debate1,dark_soliton_debate1,ds_debate2,deguchi}.
In the context of quenches, the KdV and NLSE equations have been used
for the study of dynamics of a soliton undergoing an interaction quench
\cite{GP-solitons,GP-solitons_long}.

Coming back to the problem of determining the steady state's information
content, we realise that the crucial question that arises is if the
long time asymptotics of the (operatorial now) solution of the equations
of motion involves solitary waves. In such case, memory of all initial
soliton correlations survives in the steady state. Systems exhibiting
quantum solitary waves can therefore evade decoherence and the memory
loss typically associated with the equilibration process, yet exhibiting
non-trivial dynamics without the local freezing of disorder-induced
localisation. These features make quantum solitons perfect candidates
for applications to quantum communication and quantum information
processing in many-body systems, for which reason their dynamics,
collisions and entanglement properties have been the subject of recent
studies \cite{entanglement_q-solitons,entangl,entangl_2,entangl_3,soliton_dyn,soliton_qdyn,Bell-states,coll_dyn}.
Motivated by the semiclassical approximation to the quantum KdV equation,
it is tempting to give a positive answer to the above question: the
general solution to the classical KdV equation for an arbitrary initial
condition always decomposes into a finite number of solitons and a
decaying dispersive component \cite{math:KdV_asympt}. In fact it
was exactly the observation of solitons in the classical KdV equation
by Zabusky and Kruskal that led them to the explanation of the absence
of ergodicity in the Fermi-Pasta-Ulam problem in their seminal work
\cite{ZK}. However, this semiclassical argument is insufficient in
order to address the above question in the quantum case, as it ignores
the effects of quantum fluctuations on the soliton stability.

In this work we study the large time correlation functions
of the density and velocity or phase field after a quench in a 1D
Bose gas using its quantum hydrodynamic description. We first introduce
the various effective theories (Sec.~\ref{sec:QH_models}), starting
with the Luttinger model and extending it to the quantum KdV model
by including dispersion and chiral interaction terms. We then show
that time evolution following the standard Luttinger model preserves
the memory of all initial correlation functions (Sec.~\ref{sec:LM}),
while dispersion (Sec.~\ref{sec:disp}) or interaction (Sec.~\ref{sec:int})
considered separately lead to delocalisation which suppresses this
effect: Under the condition of sufficient initial clustering, the
steady state is in both cases a Gaussian GGE in terms of the bosonisation
fields or the dual fermion field respectively. 
{The main results of our analysis are two scaling relations 
for the time decay of corrections to the steady state values of correlation functions, 
which are deduced as a combination of the scaling of the 
initial clustering and of the dynamical spreading. 
In the dispersive case such a scaling relation for exponential initial 
clustering is given in Eq.~\eqref{scaling1}, 
while in the interacting case it is expressed 
as an upper bound in Eq.~\eqref{scaling2}.} 
Next, we show that the semiclassical approximation to the quantum KdV equation suggests
that the combination of dispersion and interaction restores localisation
through the emergence of solitary waves (Sec.~\ref{sec:SC_KdV}) and,
provided that the latter persist at the quantum level, memory of soliton
correlation functions is preserved at large times (Sec.~\ref{sec:QSW}).
Lastly, we discuss the implications of our analysis (Sec.~\ref{sec:discussion}) and identify open questions for future
investigation (Sec.~\ref{sec:open_questions}).

\section{Quantum Hydrodynamics and the Quantum KdV Equation\label{sec:QH_models}}

We are interested in the dynamics of 1D non-relativistic quantum gases,
which are realised in cold atom experiments like the quasi-condensate
splitting experiments in which the first observation of a GGE was
achieved \cite{exp:GGE}. We present quantum hydrodynamics descriptions
of such systems in terms of the density and phase fields in the thermodynamic
limit. We first introduce the standard bosonisation approach \cite{Haldane}
i.e. the Luttinger model, which is essentially a free massless relativistic
field theory effectively describing the low-energy limit of such systems.
We then introduce the quantum KdV equation as a nonlinear extension
of the Luttinger model that incorporates effects of dispersion and
interactions of the bosonised fields \cite{qKdV-LL}. In particular we discuss arguments
that have been used to justify this effective description in the context
of out-of-equilibrium physics and in particular quantum quenches.

We consider an interacting non-relativistic Bose gas in 1D generally
described by the Hamiltonian

\begin{equation}
\hat{H}=\int dx\,\partial_{x}\hat{\Psi}^{\dagger}(x)\partial_{x}\hat{\Psi}(x)+\int dxdx'\,V(x-x')\hat{\Psi}^{\dagger}(x)\hat{\Psi}(x)\hat{\Psi}^{\dagger}(x')\hat{\Psi}(x')
\end{equation}
We have set the boson mass to $1/2$ and fix the number of particles
$\hat{N}=\int dx\,\hat{\Psi}^{\dagger}(x)\hat{\Psi}(x)$. In the case
of contact interaction this reduces to the integrable Lieb-Liniger
Hamiltonian 
\begin{equation}
\hat{H}_{LL}=\int dx\,\left(\partial_{x}\hat{\Psi}^{\dagger}\partial_{x}\hat{\Psi}+c\hat{\Psi}^{\dagger}\hat{\Psi}^{\dagger}\hat{\Psi}\hat{\Psi}\right)
\end{equation}
Bosonisation and more general quantum hydrodynamics arise when the
boson field operator $\hat{\Psi}$ is expressed in terms of the dual
density and phase fields $\hat{\rho}$ and $\hat{\phi}$ 
\begin{equation}
\hat{\Psi}^{\dagger}(x)=\sqrt{\hat{\rho}(x)}\mathrm{e}^{-\mathrm{i}\hat{\phi}(x)}
\end{equation}
satisfying canonical commutation relations $[\hat{\rho}(x),\hat{\phi}(x')]=\mathrm{i}\delta(x-x')$.
The original Hamiltonian $\hat{H}$ is then written as 
\begin{equation}
\hat{H}=\int dx\,\left[\hat{\rho}\left(\partial_{x}\hat{\phi}\right)^{2}+\tfrac{1}{4}\hat{\rho}^{-1}\left(\partial_{x}\hat{\rho}\right)^{2}\right]+\int dxdx'\,V(x-x')\hat{\rho}(x)\hat{\rho}(x')
\end{equation}
In order to focus on the fluctuations of the density about its mean
value $\rho_{0}=N/L$ we set, following Haldane \cite{Haldane} 
\begin{equation}
\hat{\rho}(x)=\rho_{0}+\frac{1}{\pi}\partial_{x}\hat{\theta}(x)
\end{equation}
where the new field $\hat{\theta}(x)$ satisfies the commutation relations
\begin{equation}
[\partial_{x}\hat{\theta}(x),\hat{\phi}(x')]=-[\hat{\theta}(x),\partial_{x'}\hat{\phi}(x')]=\mathrm{i}\pi\delta(x-x')
\end{equation}
In the bosonised description, local observables correspond to exponentials
of the fields $\hat{\phi}$ and $\hat{\theta}$, the ``vertex operators'',
or their derivatives. Vertex operators correspond to the original
boson field $\hat{\Psi}$, while derivatives correspond to the density
and current density operators
\begin{align}
\hat{\rho}(x) & =\hat{\Psi}^{\dagger}(x)\hat{\Psi}(x)=\rho_{0}+\frac{1}{\pi}\partial_{x}\hat{\theta}(x)\\
\hat{j}(x) & =-\frac{\mathrm{i}}{2}\left(\hat{\Psi}^{\dagger}(x)\partial_{x}\hat{\Psi}(x)-\partial_{x}\hat{\Psi}^{\dagger}(x)\hat{\Psi}(x)\right)=\sqrt{\hat{\rho}(x)}\left(\partial_{x}\hat{\phi}(x)\right)\sqrt{\hat{\rho}(x)}\sim \rho_{0}\partial_{x}\hat{\phi}(x)
\end{align}

Substituting to the above Hamiltonian and expanding in powers of $\partial_{x}\hat{\theta}$,
we obtain low energy effective Hamiltonians for the system. The simplest
low energy description is given by the standard Luttinger model Hamiltonian
\begin{equation}
\hat{H}_{Lm}=\frac{v}{2\pi}\int dx\,\left[K\left(\partial_{x}\hat{\phi}\right)^{2}+\frac{1}{K}\left(\partial_{x}\hat{\theta}\right)^{2}\right] \label{Lm1}
\end{equation}
with $v=2\rho_{0}\sqrt{\gamma}$ the sound velocity, $K=\pi/\sqrt{\gamma}$
the Luttinger liquid parameter and $\gamma=c/\rho_{0}$ the dimensionless
interaction strength.

It is convenient to introduce the left and right-moving chiral fields,
respectively $\hat{\varphi}_{+}$ and $\hat{\varphi}_{-}$, in terms
of which the Hamiltonian decouples. These are defined by 
\begin{equation}
\hat{\varphi}_{\pm}=\frac{1}{\sqrt{K}}\hat{\theta}\mp\sqrt{K}\hat{\phi} \label{chiral}
\end{equation}
and their commutation relations are therefore 
\begin{equation}
[\hat{\varphi}_{\sigma}(x),\partial_{x'}\hat{\varphi}_{\sigma'}(x')]=2\sigma\pi\mathrm{i}\delta_{\sigma\sigma'}\delta(x-x')
\label{chiral_comm-rel}
\end{equation}
We also define the rescaled bosonisation fields 
\begin{align}
\hat{\vartheta} & =\frac{1}{2}\left(\hat{\varphi}_{+}+\hat{\varphi}_{-}\right)=\frac{1}{\sqrt{K}}\hat{\theta} \label{resc} \\
\hat{\varphi} & =\frac{1}{2}\left(-\hat{\varphi}_{+}+\hat{\varphi}_{-}\right)=\sqrt{K}\hat{\phi} 
\end{align}
In terms of these fields the Luttinger model Hamiltonian reads 
\begin{equation}
\hat{H}_{Lm}=\frac{v}{4\pi}\sum_{\sigma=\pm}\int dx\,\left(\partial_{x}\hat{\varphi}_{\sigma}\right)^{2}=\frac{v}{2\pi}\int dx\,\left[\left(\partial_{x}\hat{\varphi}\right)^{2}+\left(\partial_{x}\hat{\vartheta}\right)^{2}\right] \label{Lm2}
\end{equation}

Deviations from the standard Luttinger model can be taken into account
by including higher order terms in the expansion of field derivatives.
To determine the leading corrections that are relevant at equilibrium,
one would typically apply Renormalisation Group theory arguments regarding
the relevance of perturbative corrections. While these arguments are
justified for the study of ground or thermal state physics, in the
context of quenches higher energy excitations need also to be taken
into account. This can be done to first order by including terms of
the next-to-leading scaling dimension \cite{qKdV-LL}. On the other
hand, non-chiral contributions to the Hamiltonian can be neglected
to first approximation. Such terms are expected to induce 
a modification of the velocity of the chiral excitations \cite{Cardy,Bernard-Doyon2}.

Under the above assumptions, we include only chiral terms of scaling
dimension three or four, thus obtaining the Hamiltonian 
\begin{equation}
\hat{H}=\hat{H}_{Lm}+\hat{H}_{KdV}
\end{equation}
where $\hat{H}_{KdV}$ is the quantum KdV Hamiltonian

\begin{equation}
\hat{H}_{KdV}=\frac{1}{4\pi}\sum_{\sigma=\pm}\int dx\,\left[\alpha\left(\partial_{x}\hat{\varphi}_{\sigma}\right)^{3}+\beta\left(\partial_{x}^{2}\hat{\varphi}_{\sigma}\right)^{2}\right]
\end{equation}
with $\alpha=1/(2\sqrt{K})$ and $\beta=K/(4\pi^{2}\rho_0)$. 
The first term in the above Hamiltonian
\begin{equation}
\hat{H}_{int}=\frac{\alpha}{4\pi}\sum_{\sigma=\pm}\int dx\,\left(\partial_{x}\hat{\varphi}_{\sigma}\right)^{3}
\end{equation}
is an interaction in terms of the bosonisation fields, while the second
term 
\begin{equation}
\hat{H}_{disp}=\frac{\beta}{4\pi}\sum_{\sigma=\pm}\int dx\,\left(\partial_{x}^{2}\hat{\varphi}_{\sigma}\right)^{2}
\end{equation}
is only a quadratic dispersion term.

Another convenient field transformation is introduced by the fermionisation
method \cite{refermionisation,qKdV1,BF}. Based on the boson-fermion
correspondence, introduced in the seminal works of Mattis and Mandelstam
\cite{Mandelstam,Mattis,Coleman,Mattis-Lieb,Luther-Peschel}, we define
the fermionic quasiparticle field 
\begin{equation}
\hat{\Phi}_{\sigma}^{\dagger}(x)=\hat{F}_{\sigma}\frac{1}{\sqrt{2\pi a}}\mathrm{e}^{2\pi\mathrm{i}\sigma\rho_{0}x}\mathrm{e}^{\mathrm{i}(\sigma\hat{\vartheta}(x)-\hat{\varphi}(x))}
\label{boson-fermion}
\end{equation}
where $\hat{F}_{\sigma}$ are the Klein factors and $a$ a short
distance cutoff (or lattice spacing in lattice model applications). 
Using this transformation the Luttinger model Hamiltonian can be written
as a free fermion Hamiltonian with linear dispersion
\begin{equation}
\hat{H}_{Lm}=-\mathrm{i}v\sum_{\sigma=\pm}\sigma\int dx\,:\hat{\Phi}_{\sigma}^{\dagger}(x)\partial_{x}\hat{\Phi}_{\sigma}(x):
\end{equation}
where $:\,:$ denotes normal ordering in terms of the fermions. Moreover,
the bosonic interaction term of the quantum KdV Hamiltonian can be
written as a free fermionic dispersion term \cite{qKdV1,qKdV-LL,refermionisation}
\begin{equation}
\hat{H}_{int}=\frac{1}{2m_{*}}\sum_{\sigma=\pm}\int dx\,:{\left(\partial_{x}\hat{\Phi}_{\sigma}^{\dagger}\right)\left(\partial_{x}\hat{\Phi}_{\sigma}\right)}:
\end{equation}
with $m_{*}\propto 1/\alpha$ the effective fermion mass.

We notice that the quantum KdV Hamiltonian contains only terms which
are quadratic in either the bosonised or the fermion fields \cite{qKdV1,qKdV2}.
In particular if the bosonic interaction term is absent ($\alpha=0$)
the Hamiltonian is free in terms of the bosonised fields, while if
the dispersion term is absent ($\beta=0$) the Hamiltonian is free
in terms of the fermionic field. The various Hamiltonian terms and
their physical meaning in terms of the three field theoretic descriptions,
the original boson field, the density and phase fields and the fermionic
quasiparticle field, are summarised in Table \ref{tab:LL}. 
{In particular, 
non-linear dispersion of the bosonisation fields arises as a result of an 
inter-particle interaction that is not of local (point-like) type, but non-local, 
or from the inclusion of next-to-leading order terms in the expansion of the 
kinetic energy part of the Hamiltonian. On the other hand, interaction 
of the bosonised fields arises as a result of the non-linear (non-relativistic) 
dispersion of the original bosonic particles.} More details on this 
correspondence can be found in \cite{nonlinear_LL} and a
discussion of their role on dynamics in \cite{qKdV-LL}.

\begin{table}
\begin{centering}
\renewcommand{\arraystretch}{1.5} 
\begin{tabular}{|c|c|c|c|c|}
\hline 
 original bosons/fermions $\hat{\Psi}^{\dagger}$ & \multicolumn{2}{c|}{ bosonisation fields $\hat{\phi},\hat{\theta}$ } & \multicolumn{2}{c|}{ fermionic quasiparticles $\hat{\Phi}^{\dagger}$ }\tabularnewline
\hline 
\hline 
local interaction & \multirow{3}{*}{free} & linear dispersion & free & linear disp. \tabularnewline
\cline{1-1} \cline{3-5} 
non-local interaction &  & \multirow{2}{*}{non-linear dispersion} & \multirow{3}{*}{ interacting } & \multirow{2}{*}{ interaction + nl. disp. }\tabularnewline
\cline{1-1} 
kinetic energy (next-to-leading order corrections) &  &  &  & \tabularnewline
\cline{1-3} \cline{5-5} 
\multirow{2}{*}{ non-linear dispersion / band curvature } & \multirow{2}{*}{ interacting } &{ non-chiral interaction }&  & irrelevant interaction \tabularnewline
\cline{3-5} 
 &  & chiral interaction & free & non-linear disp. \tabularnewline
\hline 
\end{tabular}
\par\end{centering}

\caption{The dictionary of Luttinger liquids.\label{tab:LL}}

\end{table}

\section{Time Evolution in the Luttinger Model\label{sec:LM}}

We derive the long time limit of correlation functions for the density
and phase fields assuming that the evolution follows the standard
Luttinger model, i.e. the free massless and relativistic boson Hamiltonian. We will see
that, due to the purely ballistic time evolution of the fields, dephasing
is suppressed and the initial correlations are preserved intact up
to infinite times, with no memory loss. The only mixing of information
that takes place is the averaging between the left and right-moving
chiral modes. Moreover tracking the initial origin of the information
that survives at large times we see that it is located at the two
spatial infinities. Therefore the large time correlations are independent
of initial correlations in any finite part in the middle. These features are 
characteristics of the linear dispersion relation
of the model.

The Heisenberg equations of motion corresponding to the standard Luttinger 
model Hamiltonian (\ref{Lm1}), (\ref{Lm2}) 
are 
\begin{align}
\partial_{t}\hat{\phi} & =-\frac{v}{K}\partial_{x}\hat{\theta}\\
\partial_{t}\hat{\theta} & =-vK\partial_{x}\hat{\phi}
\end{align}
or, in decoupled form in terms of the chiral fields (\ref{chiral}), 
\begin{equation}
\partial_{t}\hat{\varphi}_{\pm}=\pm v\partial_{x}\hat{\varphi}_{\pm}
\end{equation}
which are simply equivalent forms of the wave equation in one dimension 
$\partial_{t}^{2}\hat{\phi}=v^{2}\partial_{x}^{2}\hat{\phi}$, 
with the same equation for the field $\hat{\theta}$. Its general
solution for any initial field configuration, defined by the initial
conditions $\hat{\phi}(x,0)$ and $\partial_{t}\hat{\phi}(x,0)=-\frac{v}{K}\partial_{x}\hat{\theta}(x,0)$,
is given by d'Alembert's formula 
\begin{align}
\hat{\phi}(x,t) & =\frac{1}{2}\left(\hat{\phi}(x-vt,0)+\hat{\phi}(x+vt,0)\right)+\frac{1}{2v}\int_{x-vt}^{x+vt}dx'\,\partial_{t}\hat{\phi}(x',0)\nonumber\\
 & =\frac{1}{2}\left(\hat{\phi}(x-vt,0)+\hat{\phi}(x+vt,0)\right)+\frac{1}{2K}\left(\hat{\theta}(x-vt,0)-\hat{\theta}(x+vt,0)\right)\nonumber\\
 & =-\frac{1}{2\sqrt{K}}\sum_{\sigma=\pm}\sigma\partial_{x}\hat{\varphi}_{\sigma}(x+\sigma vt,0)
\end{align}

We are interested however in the values of \emph{local} observables
at long times, which as explained above correspond to vertex operators
or derivatives of the bosonisation fields. For simplicity here we
focus on density fluctuations, which are given in terms of the rescaled field 
(\ref{resc}) by the derivative $\partial_{x}\hat{\vartheta}$.
Its time evolution is given by 
\begin{align}
\partial_{x}\hat{\vartheta}(x,t) & =\frac{1}{2}\sum_{\sigma=\pm}\partial_{x}\hat{\varphi}_{\sigma}(x+\sigma vt,0)
\end{align}
that is, it is expressed in terms of initial local fields located
at the lightcone projection points $x\pm vt$.

\begin{figure}
\includegraphics[width=0.5\paperwidth]{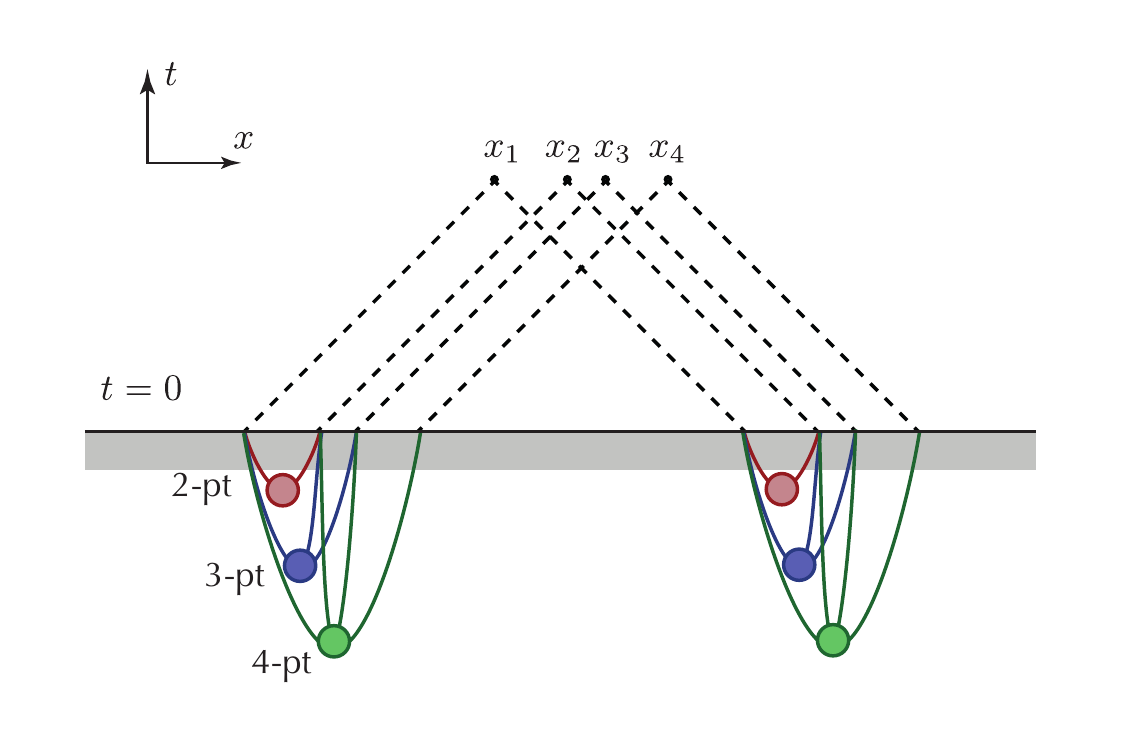}\caption{Diagrammatic representation of the calculation of long time correlations
in the standard Luttinger model. \label{fig:lightcone+cluster}}
\end{figure}

Having a direct relation between the time evolved local field
and initial local fields, it is now easy to calculate the long time
asymptotics of its correlations. The general multi-point correlation
function at time $t$ is 
\begin{align}
\left\langle \prod_{i=1}^{n}\partial_{x}\hat{\vartheta}(x_{i},t)\right\rangle 
 & =\frac{1}{2^n}\left\langle \prod_{i=1}^{n}\sum_{\sigma=\pm}\partial_{x}\hat{\varphi}_{\sigma}(x_{i}+\sigma vt,0)\right\rangle 
=\frac{1}{2^n}\sum_{\{\sigma_{i}\}}\left\langle \prod_{i=1}^{n}\partial_{x}\hat{\varphi}_{\sigma_{i}}(x_{i}+\sigma_{i}vt,0)\right\rangle 
\end{align}
where the expectation value refers to the initial state $|\Omega\rangle$.
In the long time limit the above expectation values correspond to
initial correlation functions between right chiral fields spatially
located at points tending to $-\infty$ and left chiral fields at
points tending to $+\infty$. Since these fields are local, by application
of the cluster decomposition property of the initial state we find
that as $t\to\infty$ the above correlation functions factorise to
a product of two components, one for each of the spatial infinities (Fig.~\ref{fig:lightcone+cluster})
\begin{align}
\lim_{t\to\infty}\left\langle \prod_{i=1}^{n}\partial_{x}\hat{\varphi}_{\sigma_{i}}(x_{i}+\sigma_{i}vt,0)\right\rangle & =\lim_{t\to\infty}\left\langle \prod_{i:\,\sigma_{i}=-}\partial_{x}\hat{\varphi}_{-}(x_{i}-vt,0)\right\rangle \left\langle \prod_{i:\,\sigma_{i}=+}\partial_{x}\hat{\varphi}_{+}(x_{i}+vt,0)\right\rangle \nonumber \\
& =\left\langle \prod_{i:\,\sigma_{i}=-}\partial_{x}\hat{\varphi}_{-}(x_{i},0)\right\rangle _{-\infty} \quad
\left\langle \prod_{i:\,\sigma_{i}=+}\partial_{x}\hat{\varphi}_{+}(x_{i},0)\right\rangle _{+\infty}
\end{align}
where the indices $\pm\infty$ refer to the asymptotic values of initial
correlation functions at the respective spatial infinities. 
This formula is valid even if the initial state is not translationally invariant
in the whole space but only asymptotically at the two spatial infinities
(as in the widely studied case of a system that is initially split in two 
halves at different temperature or presents any type of step-like inhomogeneity 
\cite{Bernard-Doyon}). If instead we assume that the initial state is translationally invariant,
we can drop the $\pm\infty$ indices since the expectation values at both spatial infinities are equal.

The above can be expressed elegantly in terms of \emph{connected correlation
functions}. Connected (also called `truncated') correlation functions are defined
as joint cumulants between local fields located at different points,
i.e. they are equal to correlation functions from which all possible
combinations of factorised correlations have been subtracted off.
More precisely, the above definition explains the use of the term
`truncated', while the term `connected' is justified by the linked
cluster theorem, which shows that the above defined truncated correlation
functions are identified with correlation functions defined through
connected graphs \cite{NO}. The cluster decomposition property is
rephrased as decay of the connected correlation fuctions with the
distance between any subset of points and the rest, and comes obviously
as a direct consequence of their connectedness. Expressed in momentum
space, the cluster decomposition property means that the Fourier transform
of connected correlation functions can contain at most one $\delta$-function
singularity that accounts for momentum conservation, i.e. translational
invariance of the state, but no other such singularity \cite{Weinberg}.
Milder singularities, like poles or branch cuts at the real axis are
instead allowed and reflect algebraically decaying correlations, as
in ground states of gapless systems. If there is no singularity along
the real momentum axis, but only elsewhere in the complex plane, then
correlations decay exponentially as in ground states of gapped systems
or in thermal states of quantum gases at non-zero temperature. 
A state that is Gaussian in terms of some choice of local fields, 
has vanishing connected correlation
functions for those fields for all orders higher than two. 
This is a consequence of Wick's theorem, 
which is valid precisely for Gaussian states.

From the above results, we find that connected correlation functions
between the two distinct chiral field components vanish in the large
time limit, as they originate from two different spatial infinities
at initial time. Denoting connected correlation functions by double
angular brackets $\llangle\dots\rrangle$ and using their multi-linearity,
the above result can be expressed as
\begin{align}
\lim_{t\to\infty}\llangle[\bigg]\prod_{i=1}^{n}\partial_{x}\hat{\vartheta}(x_{i},t)\rrangle[\bigg] & =\lim_{R\to\infty}\left[\llangle[\bigg]\prod_{i=1}^{n}\partial_{x}\hat{\varphi}_{-}(x_{i}-R)\rrangle[\bigg]+\llangle[\bigg]\prod_{i=1}^{n}\partial_{x}\hat{\varphi}_{+}(x_{i}+R)\rrangle[\bigg]\right]\nonumber\\
 & =\llangle[\bigg]\prod_{i=1}^{n}\partial_{x}\hat{\varphi}_{-}(x_{i})\rrangle[\bigg]_{-\infty}+\llangle[\bigg]\prod_{i=1}^{n}\partial_{x}\hat{\varphi}_{+}(x_{i})\rrangle[\bigg]_{+\infty}
\end{align}
since all other combinations of fields involve certainly both chiral
components. Equivalent results have been derived in \cite{S15} and in the more general context 
of out of equilibrium Conformal Field Theory in \cite{Bernard-Doyon1a,Bernard-Doyon1b}.

While we have focused on field derivatives, the calculation can be
easily repeated for general correlation functions of vertex operators.
As shown in \cite{S15}, in the long time limit the so-called `neutrality condition' 
is restored, which means that only correlations of chiral field \emph{differences}
between any point and some reference point survive. As for field derivatives, 
correlations between field differences originate from initial correlations
at the left and right spatial infinities, therefore the same arguments apply also here.

\section{The Effect of Nonlinear Dispersion\label{sec:disp}}

We now turn our attention to the effect of nonlinear dispersion, which 
is induced by terms in the Hamiltonian that involve higher spatial
derivatives and are quadratic in the bosonisation fields. 
Considering as before the propagation of initially localised fields, 
we realise that nonlinear dispersion leads to spreading and consequently decay
of the propagator with time. This effect, in combination with sufficient 
initial clustering of correlations, is responsible for the relaxation
towards a Gaussian GGE in terms of the bosonised fields.

The time evolution in the present case is described by the Hamiltonian
$\hat{H}_{Lm}+\hat{H}_{\text{disp}}$ with 
\begin{equation}
\hat{H}_{\text{disp}}=\frac{\beta}{2\pi}\int dx\,\left[K\left(\partial_{x}^{2}\hat{\phi}\right)^{2}+\frac{1}{K}\left(\partial_{x}^{2}\hat{\theta}\right)^{2}\right]=\frac{\beta}{4\pi}\sum_{\sigma=\pm}\int dx\,\left(\partial_{x}^{2}\hat{\varphi}_{\sigma}\right)^{2}
\end{equation}
Since this Hamiltonian is also quadratic in the bosonised fields, the time evolution
can be still derived exactly. Using the commutation relations \eqref{chiral_comm-rel}, the equations 
of motion $\partial_{t}\hat{\varphi} = i[\hat{H},\hat{\varphi}]$ can be easily shown to be
\begin{equation}
\partial_{t}\hat{\varphi}_{\pm}(x,t)=\pm\left(v\partial_{x}\hat{\varphi}_{\pm}(x,t)-\beta\partial_{x}^{3}\hat{\varphi}_{\pm}(x,t)\right)
\end{equation}
or in terms of the field $\hat{\phi}$ 
\begin{equation}
\partial^{2}_{t}\hat{\phi}(x,t)= \left(v\partial_{x}-\beta\partial_{x}^{3}\right)^{2} \hat{\phi}(x,t) \label{eom_disp}
\end{equation}
and similarly for $\hat{\theta}$. 
The general solution to the equations of motion for any quadratic bosonic 
field theory is given by
\begin{equation}
\hat{\phi}(x,t)=\int dx'\left(\partial_{t}G(x-x',t)\hat{\phi}(x',0)+G(x-x',t)\partial_{t}\hat{\phi}(x',0)\right)
\end{equation}
where $G(x,t)$ is the (retarded) Green's function of the equation
of motion, which is 
\begin{equation}
G(x,t)=\int\frac{dk}{2\pi}\,\mathrm{e}^{\mathrm{i}kx}\frac{\sin\omega(k)t}{\omega(k)} \label{Green_fn}
\end{equation}
with $\omega(k)$ the dispersion relation. 
One way to show the above is to express the equation of motion in
momentum space, solve it for arbitrary initial conditions 
\begin{equation}
\hat{\phi}_{k}(t)=\hat{\phi}_{k}(0)\cos\omega(k)t+\partial_{t}\hat{\phi}_{k}(0)\frac{\sin\omega(k)t}{\omega(k)}=\frac{1}{2}\sum_{\sigma=\pm1}\left(\hat{\phi}_{k}(0)+\sigma\frac{\partial_{t}\hat{\phi}_{k}(0)}{\mathrm{i}\omega(k)}\right)\mathrm{e}^{\mathrm{i}\sigma\omega(k)t}
\end{equation}
and then return to coordinate space.

In the present case, the dispersion relation is
\begin{equation}
\omega(k)=|k|\left(v+\beta k^{2}\right)\label{eq:disp_rel}
\end{equation}
as we can easily find by substituting the general solution 
to the equation of motion \eqref{eom_disp}. 
More generally, the phonon dispersion relation in gapless systems
is a function of the form 
\begin{equation}
\omega(k)=v|k|f(k)\sim v|k|+\lambda|k|^{\gamma}+... \label{eq:disp_rel_gen}
\end{equation}
where the function $f(k)$ satisfies $f(0)=1$ and is assumed to be
monotonically increasing, in order to ensure that bosonisation can
be applied \cite{Haldane}. In the present case the first order correction
to the linear dispersion is cubic i.e. $\gamma=3$ with $\lambda>0$.
In non-relativistic interacting Bose gases the phonon dispersion is typically of
the Bogoliubov form, whose leading low-momentum correction to the
linear dispersion is precisely as in \eqref{eq:disp_rel}. However
there is no fundamental reason preventing a quadratic correction i.e.
$\gamma=2$: in fact this type of dispersion is encountered in the
Calogero-Sutherland model whose quantum hydrodynamic field theory
description is given by the Benjamin-Ono equation \cite{q_hydrodyn2,B-O2}.
In gapless lattice models with finite range interactions, like the
Ising model in transverse field, XY or XXZ models in their critical
phases, we find cubic corrections with $\lambda<0$ consistently 
with the Lieb-Robinson bound.

For a linear dispersion relation, we set $\beta=0$ and the Green's function is
\begin{equation}
G(x,t)=\int\frac{dk}{2\pi}\,\mathrm{e}^{\mathrm{i}kx}\frac{\sin v|k|t}{v|k|}=\frac{1}{2v}\Theta(vt-|x|)
\end{equation}
thus we recover d'Alembert's formula. For non-linear dispersion the
Green's function retains qualitatively its step-like form, but it is no longer sharp: 
it is smeared over a length that grows algebraically with time. This
means in particular that field derivatives, which are those that represent local fields, 
are no longer simply equal to initial fields at the two light-cone projection points $x\pm vt$,
but instead convolutions of initial derivative fields, centered at
those points, which now spread and decay with time. More explicitly we have
\begin{align}
\partial_{x}\hat{\vartheta}(x,t) & =\int dx'\left(\partial_{t}G(x-x',t)\partial_{x'}\hat{\vartheta}(x',0)+\partial_{x}G(x-x',t)\partial_{t}\hat{\vartheta}(x',0)\right)\nonumber\\
 & =\sum_{\nu=0,1}\int dx'\,G^{(\nu,1-\nu)}(x-x',t)\hat{\vartheta}^{(1-\nu,\nu)}(x',0) \label{disp_theta}
\end{align}
with the shorthand notation $f^{(\nu,\mu)}(x,t)\equiv\partial_{x}^{\nu}\partial_{t}^{\mu}f(x,t)$.
From the expression \eqref{Green_fn} for the Green's function $G(x,t)$, we can calculate
its derivatives that enter in the last expression 
\begin{align}
\partial_{t}G(x,t) & =\int\frac{dk}{2\pi}\,\mathrm{e}^{\mathrm{i}kx}\cos\omega(k)t=\sum_{\sigma=\pm}\Delta^{(0)}(x+\sigma vt;\sigma t) \label{dtG} \\
\partial_{x}G(x,t) & =\int\frac{dk}{2\pi}\,\mathrm{e}^{\mathrm{i}kx}\frac{\mathrm{i}k}{\omega(k)}\sin\omega(k)t=\sum_{\sigma=\pm}\sigma\Delta^{(1)}(x+\sigma vt;\sigma t) \label{dxG}
\end{align}
where we have defined the functions 
\begin{align}
\Delta^{(0)}(r;t) & =\frac{1}{2}\int\frac{dk}{2\pi}\,\mathrm{e}^{\mathrm{i}kr}\mathrm{e}^{\mathrm{i}vk(f(k)-1)t}\\
\Delta^{(1)}(r;t) & =\frac{1}{2}\int\frac{dk}{2\pi}\,\mathrm{e}^{\mathrm{i}kr}\frac{\mathrm{e}^{\mathrm{i}vk(f(k)-1)t}}{vf(k)}
\end{align}
with $f(k)$ given by \eqref{eq:disp_rel_gen}. 
(Notice that since the functions $\cos x$ and $(\sin x)/x$ are even, 
we can remove the absolute value that appears in \eqref{eq:disp_rel_gen} 
from the expressions \eqref{dtG} and \eqref{dxG} above, which we 
have used in the definitions of the functions $\Delta^{(0)}$ and $\Delta^{(1)}$.)  
The above functions play the role of propagators of local fields. 
Using the general form of the function $f(k)$ as defined in 
\eqref{eq:disp_rel_gen}, we observe
that both of these functions are centered at $r=0$, so that in \eqref{dtG} 
and \eqref{dxG} they follow the light-cone projection points $x\pm vt$. 
Moreover we find that they spread with time and their central values 
decay as $\sim t^{-1/\gamma}$. 
In the special case of cubic dispersive correction \eqref{eq:disp_rel}
considered here, $\Delta^{(0)}(r;t)=\frac{1}{2}\text{Ai}\left(r/(3\beta t)^{1/3}\right)/(3\beta t)^{1/3}$
where $\text{Ai}(x)$ denotes the Airy function.

\begin{figure}
\includegraphics[width=0.45\paperwidth]{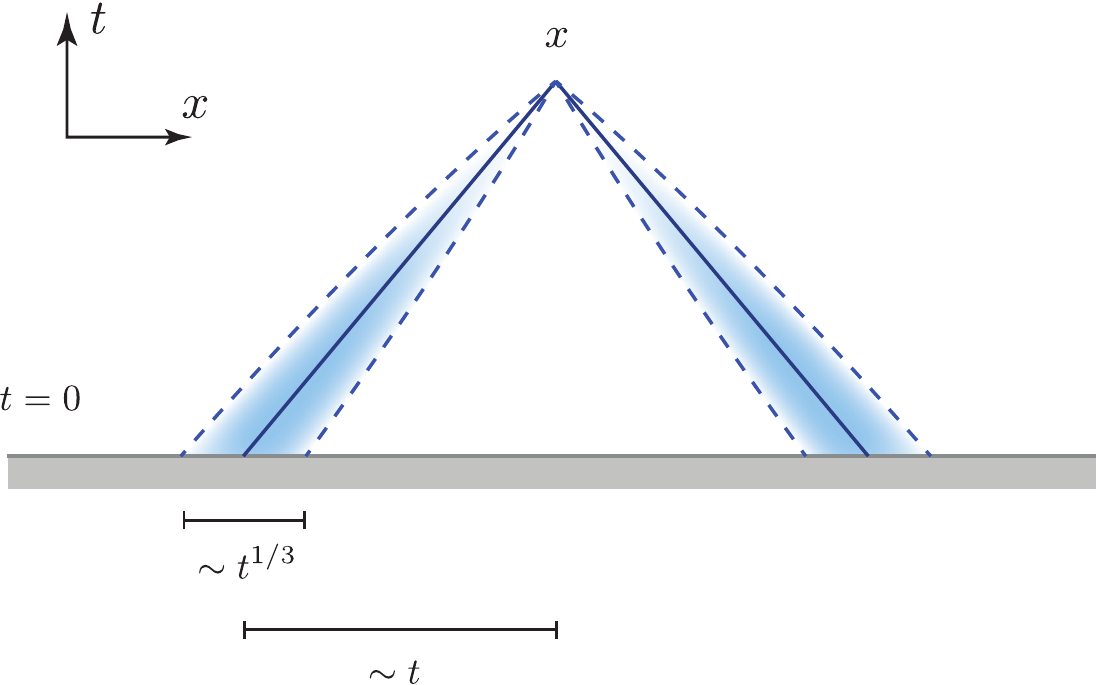}\caption{The effect 
of nonlinear dispersion is to blur the light-cone projections 
from precise points to spatial intervals, which however
remain well separated since their width scales slower with time ($\sim t^{1/3}$ for cubic nonlinearity)
than their spatial separation ($\sim t$). \label{fig:dispersion}}
\end{figure}

Substituting in the expression for the time evolution of $\partial_{x}\hat{\vartheta}(x,t)$ \eqref{disp_theta}
we find
\begin{align}
\partial_{x}\hat{\vartheta}(x,t)  =\sum_{\sigma=\pm1}\sum_{\nu=0,1}\sigma^{\nu}\int dx'\,\Delta^{(\nu)}(x-x'+\sigma vt;\sigma t)\hat{\vartheta}^{(1-\nu,\nu)}(x',0)
\end{align}
Note that this is a relation between time evolved local fields and
initial local fields. As anticipated, from the expressions for the
propagators $\Delta^{(\nu)}(r;t)$ we see that the above convolutions
are still centered at the points $x\pm vt$, but their width spreads
with time and their central peak decays with time (Fig. \ref{fig:dispersion}).
In particular, their time decay is uniform in space, i.e. at any time
$t$ they satisfy the relation $\left|\Delta^{(\nu)}(r;t)\right|<\text{(const.)}\times t^{-1/\gamma}$
for all $r$. The same spreading effect has been observed in the dynamics of 
perturbed Conformal Field Theories \cite{Bernard-Doyon2,Cardy} and in the original 
Tomonaga-Luttinger model with non-local interactions \cite{LLMM}, which also 
correspond to nonlinear dispersion in terms of the bosonisation fields (Table~\ref{tab:LL}).

From the explicit relation between time evolved local fields and initial local fields, 
we can easily derive the long time asymptotics of connected correlation functions. 
We find that, as a result of the nonlinear dispersion,
they vanish at large times for all orders larger than two, i.e. the
large time limit is described by a Gaussian state in terms of the
bosonisation fields. We will report our results below and present an intuitive explanation, 
leaving the rigorous derivation for App.~\ref{app:dispersion}. 

Let us start with the two-point function, for which we obtain \eqref{disp_2pt-cf_a}
\begin{align}
\lim_{t\to\infty}\llangle[\bigg]\partial_{x}\hat{\vartheta}(x,t)\partial_{y}\hat{\vartheta}(y,t)\rrangle[\bigg] & =\frac{1}{2}\int\frac{dk}{2\pi}\,\mathrm{e}^{\mathrm{i}k(x-y)}\left(\tilde{C}_{0}^{(0,0)}(k)+\frac{1}{v^{2}f^{2}(k)}\tilde{C}_{0}^{(1,1)}(k)\right) \label{disp_2pt-cf}
\end{align}
Note that the large time two-point correlation function depends significantly on the nonlinearity of the dispersion relation through the function $f(k)$, even if this dependence does not affect its large distance asymptotics, since $f(k)$ is smooth for $k\to 0$ with $f(0)=1$. 

For higher order connected correlation functions $n>2$, 
we find that they decay algebraically with time and the scaling of decaying corrections is controlled
by both the large distance decay of initial correlations and the leading
power law correction to the linear dispersion relation, i.e. by the
low energy limit of 
\begin{equation}
\omega(k)-v|k|\sim\lambda|k|^{\gamma}
\end{equation}
according to \eqref{eq:disp_rel_gen}. Assuming exponential initial clustering, 
in which case connected correlation functions have no singularity
for real small momenta, we obtain the general scaling law \eqref{scaling1_a}
\begin{equation}
\lim_{t\to\infty}\llangle[\bigg]\prod_{i=1}^{n}\partial_{x}\hat{\vartheta}(x_{i},t)\rrangle[\bigg]\sim\frac{1}{t^{(n-1)/\gamma}}F(\{x_{i}/t^{1/\gamma}\})
\label{scaling1}
\end{equation}
where $F(\{s_{i}\})$ has a finite non-zero value when all of its arguments
are zero. For algebraic initial clustering, additional multiplicative
factors are expected, depending on the low-momentum behaviour of initial
correlations. Overall this suggests a certain universality of the
scaling of time decaying corrections. Note that in the present case
as well as for a Bogoliubov-type dispersion relation, $\gamma=3$
and $\lambda=\frac{1}{2}v\beta$. The same exponent $\gamma=3$ appears
in the dispersion relation of typical critical lattice systems, like
the critical Ising model, and was proposed to control in a universal way 
the perturbative corrections to Conformal Field Theory predictions 
in quantum transport \cite{Bernard-Doyon2}. The time decay
of transients in out-of-equilibrium problems has been studied extensively
both in quench and transport problems \cite{long-t-cor_3,long-t-cor_2,long-t-cor_4,
long-t-corrections_1,bertini_long-t-cor,Bernard-Doyon},
revealing a certain universality of the scaling laws, in agreement
with our analysis.

The vanishing of connected correlation functions of order $n>2$ at large times can
be seen as a direct consequence of initial clustering and of the
uniform time decay of the propagators due to dispersion, using the following {simplified and intuitive
(though non-rigorous) argument}: For sufficient initial
clustering, such that the spatial integral of the initial correlation
function over the coordinate differences in the whole space is convergent,
the large time limit is bounded by the time decaying factors contributed
by all but one propagators. In more detail, the relation of the
time evolved correlations in terms of initial ones can be written as
\begin{equation}
C(x_{1},...,x_{2};t)=\int_{-\infty}^{+\infty}\prod_{i=1}^{n}dx'_{i}\,\prod_{i=1}^{n}\,G(x_{i}-x'_{i};t)C_{0}(x'_{1},x'_{2},...,x'_{n})
\end{equation}
\begin{figure}
\includegraphics[width=0.3\paperwidth]{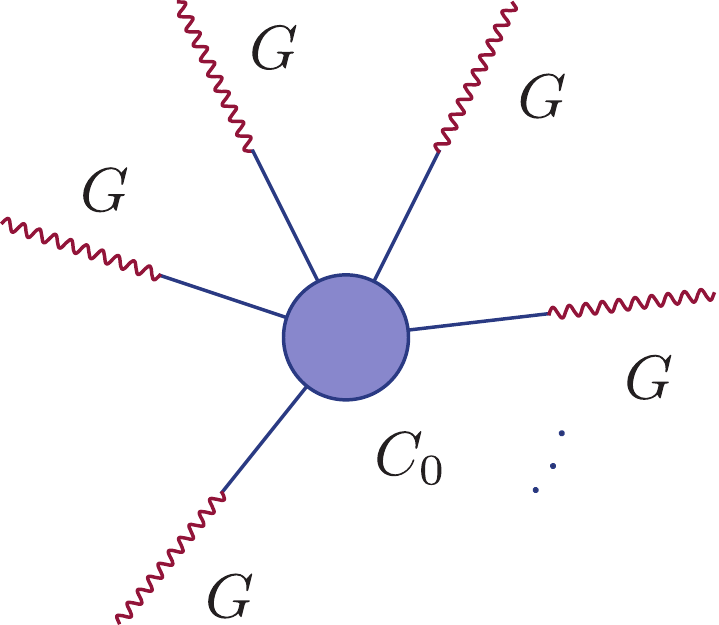}\caption{General form of Feynman diagrams for the time evolved connected correlation
functions after a quantum quench from an interacting to a free field
theory. The central graph (in blue) corresponds to the initial connected
correlation function $C_{0}$, while the attached extensions of the
external legs (curly red lines) correspond to the propagators $G$
that perform the time evolution. \label{fig:feynman_diagr}}
\end{figure}
In the diagrammatic language of quantum field theory, this expression 
means that time evolution amounts to extending the external legs of
initial correlation functions by free propagators of the post-quench
Hamiltonian (Fig. \ref{fig:feynman_diagr}). 
Assuming initial clustering such that 
\begin{equation}
\int_{-\infty}^{+\infty}\prod_{i=2}^{n}dr_{i}\,\left|C_{0}(0,r{}_{2},...,r_{n})\right|=c_{0}<\infty \label{cond1}
\end{equation}
and propagator uniformly decaying with time
\begin{equation}
\left|G(x;t)\right|<g(t)\to0,\text{ as }t\to\infty \label{cond2}
\end{equation}
with spatial integral bounded as
\begin{equation}
\int_{-\infty}^{+\infty}dr\,G(r;t)<h \label{cond3}
\end{equation}
the following bound on the time evolved correlations holds 
\begin{align}
\left|C(x_{1},...,x_{n};t)\right| & \leq\left|\int_{-\infty}^{+\infty}dx'_{1}\,G(x_{1}-x'_{1};t)\int_{-\infty}^{+\infty}\prod_{i=2}^{n}d(x'_{i}-x'_{1})\,\prod_{i=2}^{n}\left|G(x_{i}-x'_{i};t)\right|\left|C_{0}(0,x'_{2}-x'_{1},...,x'_{n}-x'_{1})\right|\right|\nonumber\\
 & <c_{0}h\left(g(t)\right)^{n-1}\to0,\text{ as }t\to\infty\quad(n>2)\label{bound}
\end{align}
In the above we assumed a translationally invariant initial state
and dropped the field component indices $\nu_i$ to lighten the notation. 
The condition \eqref{cond1} on the finiteness of the spatial integral of 
initial correlations should be read as 
``finite in the thermodynamic limit'', i.e. exhibiting no infrared divergence 
(ultraviolet divergences are irrelevant in this physical context). 
The condition \eqref{cond3} of bounded spatial integral of the propagator is generally 
satisfied as a result of conservation of particle number, which in fact means that  $\int_{-\infty}^{+\infty}dr\,G(r;t)=\text{const.}$.
Note also that in the above argument, a uniform decay of the propagator
is required, i.e. at any time it must be bounded in the whole space
by a time decaying function \eqref{cond2}, otherwise initial fluctuations may be
transported away from their original spatial point but without decaying.
We can easily see now the connection between dispersion and time decay
of correlations: in systems with conserved particle number, the integral
of the propagator over all space is constant in time and therefore
dispersive spreading of the propagator is related to decay with time
of its maximum value as a function of the spatial coordinate.

{Even though the above argument is quite intuitive,
it is not rigorous, as the validity of the inequalities relies on mathematical conditions 
that are not strictly satisfied for the involved functions. For this reason 
cancellations of oscillatory terms are disregarded by this argument. That is why the argument fails for $n=2$:
in this case, the two-point correlations involve terms in which the time dependent oscillatory
terms exactly cancel due to the condition of momentum conservation, thus giving a non-vanishing contribution at large times. 
In contrast, for $n>2$ the momentum conservation condition is not sufficient to ensure the cancellation of oscillatory
terms (such cancellations occur only in regions of zero integration measure and do not contribute in the large time limit). 
A rigorous derivation of the large time asymptotics is given in App.~\ref{app:dispersion}.}

As a last comment, we mention that the monotonicity of the dispersion relation $\omega(k)$
is a non-trivial and relevant assumption, as it affects the form of
the Fourier transforms of the derivatives of the propagator. In the
context of non-relativistic quantum gases, this assumption breaks
down at the bottom of the Luttinger liquid Fermi sea. This is expected
to lead to additional time decaying factors that are typically captured
by the ``mobile impurity'' approach of nonlinear Luttinger liquids
\cite{nonlinear_LL}.

\section{The Effect of Chiral Interaction\label{sec:int}}

We now focus on the opposite limit, in which interaction is present but no dispersion. 
Using the boson-fermion correspondence, we will see that
the bosonic interaction is equivalent to dispersion of the fermionic
quasiparticles, thus leading again to a Gaussian GGE but in terms
of the fermionised fields this time.

The time evolution is described by the Hamiltonian $\hat{H}_{Lm}+\hat{H}_{int}$
with 
\begin{equation}
\hat{H}_{int}=\frac{\alpha}{4\pi}\sum_{\sigma=\pm}\int dx\,\left(\partial_{x}\hat{\varphi}_{\sigma}\right)^{3}
\end{equation}
Despite the presence of bosonic interaction, this Hamiltonian can
be solved exactly by means of the fermionisation method \cite{BF}. 
Indeed, as shown in \cite{refermionisation,qKdV1}, when expressed 
in terms of the fermionic quasiparticle field defined through 
the bosonisation formula \eqref{boson-fermion}
\begin{equation}
\hat{\Phi}_{\sigma}^{\dagger}(x)=\hat{F}_{\sigma}\frac{1}{\sqrt{2\pi a}}\mathrm{e}^{2\pi\mathrm{i}\sigma\rho_{0}x}\mathrm{e}^{\mathrm{i}(\sigma\hat{\vartheta}(x)-\hat{\varphi}(x))} \label{boson-fermion2}
\end{equation}
the Hamiltonian $\hat{H}_{int}$ reduces to the non-interacting Hamiltonian of non-relativistic
fermions 
\begin{equation}
\hat{H}_{int}=\frac{1}{2m^*}\sum_{\sigma=\pm}\int dx\,:\left(\partial_{x}\hat{\Phi}_{\sigma}^{\dagger}\right)\left(\partial_{x}\hat{\Phi}_{\sigma}\right):
\end{equation}
with effective mass $m^*\propto 1/\alpha$.

We will study the time evolution and large time asymptotics of
fermionic correlation functions, focusing in particular on density correlations. 
The method we use is analogous to 
that of non-relativistic bosons studied in \cite{SC_GGE-val}
and that of the Tonks-Girardeau dynamics of a BEC initial state \cite{KCC} 
which in some sense we extend to more general initial states, relying merely on the 
assumption of clustering of initial fermionic correlations. 
Indeed as we will see, for the same reasons as in the dispersive case, the large time correlation 
functions decompose into two contributions, one from each of the 
two spatial infinities corresponding to the two decoupled chiral fields $\sigma=\pm1$. 
We can therefore focus on one of the two chiral components, 
removing the ballistic coordinate shift of the propagator. The physical problem
is then mathematically equivalent to the quench dynamics 
under the Tonks-Girardeau Hamiltonian, 
starting from a general initial state. 

The Heisenberg equation of motion for the fermion field is 
\begin{align}
\mathrm{i} \partial_{t} \hat{\Phi}_{\pm}(x,t) & =\pm\mathrm{i}v\partial_{x}\hat{\Phi}_{\pm}(x,t) 
-\frac{1}{2m^*}\partial^{2}_{x}\hat{\Phi}_{\pm}(x,t)
\end{align}
with solution 
\begin{equation}
\hat{\Phi}_{\sigma}(x,t)=\int\mathrm{d}x'G_{f}^{\sigma}(x-x',t)\hat{\Phi}_{\sigma}(x',0)
\end{equation}
where 
\begin{equation}
G_{f}^{\sigma}(x-x',t)=\int\frac{\mathrm{d}k}{2\pi}\,\mathrm{e}^{\mathrm{i}k(x-x')+\mathrm{i}k\sigma vt-\mathrm{i} {k^{2}}t/(2m^*)}=\sqrt{\frac{m^*}{2\pi\mathrm{i} t}}\mathrm{e}^{\mathrm{i}m^*(x-x'+\sigma vt)^{2}/(2t)}\label{eq:int_prop}
\end{equation}

The fermionic two-point correlation function is time independent
and automatically described by the fermionic Gaussian GGE, since it
is directly related by Fourier transform to the conserved momentum-mode
occupation number operators $\hat{n}_{\sigma}(k)=\hat{\Phi}_{\sigma k}^{\dagger}\hat{\Phi}_{\sigma k}$,
which play the role of conserved charges of the GGE \cite{SC_GGE-val,KCC}.
The first non-trivial check of Gaussianity of the steady state
comes from the study of density-density correlations and higher order
ones. As before we will restrict ourselves to density correlations. 

From the bosonisation formula \eqref{boson-fermion2} 
the density field can be easily expressed in terms 
of the densities of fermionic fields
\begin{align}
\delta\hat{\rho}(x)&\equiv\hat{\Psi}^{\dagger}(x)\hat{\Psi}(x)-\rho_0 =\frac{1}{\pi}\partial_{x}\hat{\theta}(x) = \sum_{\sigma=\pm}\hat{\Phi}_\sigma^{\dagger}(x)\hat{\Phi}_\sigma(x)
\end{align}
From the above we find that the time evolution of the density 
operator is given in terms of the fermion fields by
\begin{align}
\delta\hat{\rho}(x,t) & =\sum_{\sigma=\pm}\int\mathrm{d}s\mathrm{d}s'\,G_{f}^{\sigma*}(x-s,t)G_{f}^\sigma(x-s',t)\;\hat{\Phi}_\sigma^{\dagger}(s,0)\hat{\Phi}_\sigma(s',0)
\end{align}
This formula for the general solution 
of the equations of motion has been derived in equivalent 
form in \cite{qKdV1,qKdV2}. 
We therefore find that the $n$-point density correlation function 
is given in terms of initial $2n$-point fermionic correlation functions
\begin{align}
\left\langle \prod_{i=1}^{n}\delta\hat{\rho}(x_{i},t)\right\rangle  & =\sum_{\sigma_{i}=\pm}\int\mathrm{d}s_{i}\mathrm{d}s_{i}'\,\prod_{i=1}^{n}G_{f}^{\sigma_{i}*}(x_{i}-s_{i},t)G_{f}^{\sigma_{i}}(x_{i}-s_{i}',t)\;\left\langle \prod_{i=1}^{n}\hat{\Phi}_{\sigma_{i}}^{\dagger}(s_{i},0)\hat{\Phi}_{\sigma_{i}}(s'_{i},0)\right\rangle 
\end{align}
For the connected correlation functions of the density, using the multi-linearity of connected correlations, we find
\begin{align}
&\llangle[\bigg]\prod_{i=1}^{n}\delta\hat{\rho}(x_{i},t)\rrangle[\bigg] \nonumber \nonumber 
=\sum_{\sigma_{i},\sigma'_{i}=\pm}\int\mathrm{d}s_{i}\mathrm{d}s_{i}'\,\prod_{i=1}^{n}G_{f}^{\sigma_{i}*}(x_{i}-s_{i},t)G_{f}^{\sigma'_{i}}(x_{i}-s_{i}',t)\;C_{0f}^{\{\sigma_{i},\sigma'_{i}\}}\left(\{(s_{i},s_{i}')\}\right)\nonumber \nonumber\\
 & =\sum_{\sigma_{i},\sigma'_{i}=\pm}\int\prod_{i=1}^{n}\frac{dk_{i}dk'_{i}}{(2\pi)^{2}}\,\mathrm{e}^{\mathrm{i}\sum_{i}\left[-(k_{i}-k'_{i})x_{i}-\mathrm{i}\left(k_{i}\sigma_{i}-k'_{i}\sigma'_{i}\right)vt+\mathrm{i}(k_{i}^{2}-\left.k'_{i}\right.^{2}) t/(2m^*)\right]}\;\delta({\textstyle \sum_{i}k_{i}-\sum_{i}k'_{i}})\tilde{C}_{0f}^{\{\sigma_{i},\sigma'_{i}\}}\left(\{(k_{i},k'_{i})\}\right)\label{eq:int_correl}
\end{align}
where 
\begin{equation}
C_{0f}^{\{\sigma_{i},\sigma'_{i}\}}\left(\{(s_{i},s_{i}')\}\right)=\llangle[\bigg]\prod_{i=1}^{n}\hat{\Phi}_{\sigma_{i}}^{\dagger}(s_{i},0)\hat{\Phi}_{\sigma_{i}}(s'_{i},0)\rrangle[\bigg]
\end{equation}
are $2n$-point initial fermionic connected correlation functions. 

At this point we should analyse the condition of cluster decomposition property of initial fermionic correlation functions, since there are some obvious differences with the previous case. This condition is non-trivial because, unlike the boson field $\hat{\Psi}^{\dagger}(x)$, the fermion field $\hat{\Phi}_{\sigma}^{\dagger}(x)$ is essentially non-local. Indeed from the bosonisation formula \eqref{boson-fermion2} which defines 
the fermion field in terms of the bosonisation fields, by expressing 
the exponent in terms of chiral field derivatives, which are local fields, we find
\begin{equation}
\hat{\Phi}_{\sigma}^{\dagger}(x) \sim \mathrm{e}^{\mathrm{i}\pi\int_{-\infty}^{x}\hat{\rho}_\sigma(x') \mathrm{d}x'}
\label{JW}
\end{equation}
where $\hat{\rho}_\sigma(x)=\partial_{x}\hat{\varphi}_\sigma(x)/\pi$ 
corresponds to densities of chiral fields. 
The above relation is essentially a Jordan-Wigner type of 
transformation \cite{nonlinear_LL}, which is manifestly non-local. {Despite this fact, it can be shown that at least a weak version of the cluster property is still valid for the fermionic correlations. While we cannot provide a general proof that the strong version of cluster property is also valid, i.e. that clustering of bosonic correlations would also imply clustering of fermionic correlations, nevertheless by studying special types of fermionic correlation functions we find evidence that this may be true. In fact even if bosonic correlations exhibit mild algebraic clustering, fermionic correlations can exhibit exponential clustering, as is the case for at least one type of state studied in the literature \cite{KCC}. A detailed analysis of fermionic clustering property is given in App.~\ref{app:fermionic_clustering}.}

Assuming that the clustering property holds for the fermionic correlations, 
we can follow the same arguments as in the previous dispersive case. 
Since the propagator \eqref{eq:int_prop}
decays with time uniformly in the whole space, assuming sufficiently
fast decaying initial correlations, the connected correlations decay
with time. We also find that the leading contributions to the large
time limit come from the left and right spatial infinities, since
the propagators spread slower than the ballistic separation of the
left and right lightcone projections. This can be seen from \eqref{eq:int_correl}:
it is only the terms with all $\sigma_{i}=\sigma'_{i}$ that
determine the slowest time decay, since all other terms come with
extra factors of $1/t$. Therefore
\begin{align}
\lim_{t\to\infty}\llangle[\bigg]\prod_{i=1}^{n}\delta\hat{\rho}(x_{i},t)\rrangle[\bigg] & =\lim_{t\to\infty}\sum_{\sigma=\pm}\int\prod_{i=1}^{n}\frac{dk_{i}dk'_{i}}{(2\pi)^{2}}\,\mathrm{e}^{-\mathrm{i}\sum_{i}(k_{i}-k'_{i})x_{i}+\mathrm{i}\sum_{i}(k_{i}^{2}-\left.k'_{i}\right.^{2}) t/(2m^*)}\;\delta({\textstyle \sum_{i}k_{i}-\sum_{i}k'_{i}})\tilde{C}_{0f}^{\sigma}\left(\{(k_{i},k'_{i})\}\right)
\end{align}
By re-scaling the integration variables, the above formula suggests
as a maximum bound of the time decay the following scaling relation
\begin{equation}
\lim_{t\to\infty}\llangle[\bigg]\prod_{i=1}^{n}\delta\hat{\rho}(x_{i},t)\rrangle[\bigg]<t^{-(2n-1)/2}F_f(\{x_{i}^{2}/t\})
\label{scaling2}
\end{equation}
However the time decaying factor in this relation is overestimated,
since the antisymmetry of the initial fermionic correlation functions
means that there may be extra factors of $k_i,k'_i$ in the integrand that downgrade it further.

To summarise this section, we have shown that, provided that initial correlations 
satisfy sufficient clustering (in terms of the fermion field this time), 
connected correlation functions of the density of order $n>1$ decay with time, as
a consequence of the time decay of the propagators. This means that
the large time limit of density correlations is given by an ensemble that is 
Gaussian in terms of the fermionic quasiparticle field.

\section{Semiclassical Analysis of the Quantum KdV Equation\label{sec:SC_KdV}}

We now consider the time evolution under the quantum KdV Hamiltonian.
Integrability of the quantum KdV and related modified equations has been
studied due to their connection with Conformal Field Theories \cite{BLZ}.
The construction of commuting local conserved charges has been studied
in \cite{qKdV_charges} and the Algebraic Bethe Ansatz formulation
has been studied in \cite{KdV+ABA}. Unlike the previous essentially
free models, the exact general solution of the Heisenberg equations
of motion is not known, however some hints about the nature of the dynamics can
be drawn from the semiclassical approximation \cite{qKdV-LL,GP-solitons}.
The latter ignores the effects of quantum fluctuations on local fields,
but is expected to give qualitatively correct results. Under this
approximation, the time evolution of a local perturbation of the density
is described by the classical KdV equation, which is known to exhibit
solitary wave solutions, i.e. ballistically moving localised and shape-preserving
waves. Their existence, perhaps the most characteristic feature of
integrability, is due to a cancellation of the effects of dispersion
and non-linearity (interaction). Moreover it is known that the long
time asymptotics of the solution for a general initial condition consists
always of a set of solitary waves with different velocities and a continuum
of decaying dispersive waves \cite{math:KdV_asympt}. In particular
point-like initial fluctuations always decompose asymptotically to
exactly one solitary wave for each of the two chiral modes, together
with a dispersive component. Below we review the semiclassical analysis
as well as general results for the long time asymptotics of the classical
KdV in more detail.

We first derive the equations of motion corresponding to the quantum KdV Hamiltonian 
\begin{equation}
\hat{H}_{KdV}=\frac{1}{4\pi}\sum_{\sigma=\pm}\int dx\,\left[\alpha\left(\partial_{x}\hat{\varphi}_{\sigma}\right)^{3}+\beta\left(\partial_{x}^{2}\hat{\varphi}_{\sigma}\right)^{2}\right]
\end{equation}
As in the classical KdV equation, the parameters $\alpha$ and $\beta$ can 
be absorbed by a suitable redefinition of space and time variables. We can therefore write
\begin{equation}
\sigma\partial_{t}\hat{\varphi}_{\sigma}=\frac{3}{2}:\left(\partial_{x}\hat{\varphi}_{\sigma}\right)^{2}:-\partial_{x}^{3}\hat{\varphi}_{\sigma}
\end{equation}
Since the two chiral components of the field evolve independently,
we can focus on one of them, say the right moving one, $\sigma=+$. The
solution for the other chiral component can be found by a simple space
reflection transformation. The equation for the local derivative field
$\hat{u}=\frac{1}{2}\partial_{x}\hat{\varphi}_{+}$ is 
\begin{equation}
\partial_{t}\hat{u}=3:\left[\left(\partial_{x}\hat{u}\right)\hat{u}+\hat{u}\left(\partial_{x}\hat{u}\right)\right]:-\partial_{x}^{3}\hat{u}
\end{equation}

We are interested in the localisation properties of the operator-valued
field solution of the above equation in the long time limit. In absence
of exact results for the general solution, we restrict ourselves to
a partial check that can be performed using the semiclassical approximation.
More specifically we will consider the dynamical response of the system
to an arbitrary localised initial fluctuation, i.e. a local quantum
quench. This problem has been previously studied in \cite{GP-solitons}
considering as initial state a soliton configuration corresponding
to different values of the parameters of the Hamiltonian. It was shown
that this pre-quench soliton splits into two stable localised wavepackets
moving to opposite directions. This should be compared with the previously
studied cases: in the presence of either dispersion or interaction
the dynamics for such a local quench would result in spreading of
the initially localised fluctuation and decay of its amplitude with time.

The semiclassical approximation corresponds to considering the expectation
value of the above equation in some state of the bosonised Hilbert
space and ignoring the effects of local quantum fluctuations, i.e.
approximating the expectation value of the product of operators in
the above equation by the product of their expectation values. Setting
$u(x,t)=\langle\hat{u}(x,t)\rangle$ and applying the semiclassical
approximation in the above equation for the right moving chiral mode,
we end up with the standard form of the classical KdV equation

\begin{equation}
\partial_{t}u(x,t)=6u(x,t)\partial_{x}u(x,t)-\partial_{x}^{3}u(x,t)
\end{equation}
As is well-known, this classical equation admits solitary wave solutions, 
as well as multi-soliton solutions and dispersive wave solutions. Its 
general solution for an arbitrary initial condition is given by the inverse scattering
transform. In combination with the nonlinear steepest descent method, 
it has be shown that at long times any initially localised field profile decomposes to
a superposition of one or more solitons moving at different
velocities to the same direction and a decaying in time dispersive
part moving to the opposite direction \cite{math:KdV_asympt}. 
{A short overview of the properties of the classical KdV equation 
is given in App.~\ref{app:classical_KdV}.} The above general result shows that, at least in the regime of validity
of the semiclassical approximation, localised field fluctuations evolve
so that they always form ballistically moving localised excitations.
The effect of quantum fluctuations may however affect 
the stability of solitons, ultimately leading to complete
decay of these excitations. Below we analyse the effects of potentially 
stable quantum solitons on equilibration.

\section{Quantum Solitary Waves and Equilibration\label{sec:QSW}}

We showed that evolution under the standard Luttinger model preserves
the memory of initial correlations up to infinite times. We will now
show that this effect can be interpreted as a consequence of the ballistic and localised
character of excitations in the standard Luttinger model. Motivated
by the semiclassical analysis of the quantum KdV equation, we then
conclude that whenever solitary waves remain stable at the quantum level, 
equilibration exhibits the same memory preservation effect.

As explained earlier, using the boson-fermion correspondence 
\begin{equation}
\hat{\Phi}_{\sigma}^{\dagger}(x)=\hat{F}_{\sigma}\frac{1}{\sqrt{2\pi a}}\mathrm{e}^{2\pi\mathrm{i}\sigma\rho_{0}x}\mathrm{e}^{\mathrm{i}(\sigma\hat{\vartheta}(x)-\hat{\varphi}(x))}
\end{equation}
the Luttinger model can be equivalently written as a system of free
fermions with linear dispersion, i.e. free massless Majorana fermions
in the language of relativistic field theory

\begin{align}
\hat{H}_{Lm} & =-\mathrm{i}v\sum_{\sigma=\pm}\sigma\int dx\,:\hat{\Phi}_{\sigma}^{\dagger}(x)\partial_{x}\hat{\Phi}_{\sigma}(x):
\end{align}
The operators $\hat{\Phi}_{\sigma}^{\dagger}(x)$ can be recognised
as soliton creation operators \cite{Mandelstam}, since their commutation relations with
the bosonisation field $\hat{\varphi}(x)$ are
\begin{align}
[\hat{\varphi}_{\sigma}(x),\hat{\Phi}_{\sigma}^{\dagger}(x')] & =\pi\Theta(x-x')\hat{\Phi}_{\sigma}^{\dagger}(x')
\end{align}
where $\Theta(x)$ is the step function. This means that $\hat{\Phi}_{\sigma}^{\dagger}(x)$
creates a static soliton state out of the vacuum. The equations of
motion for the soliton operator under $\hat{H}_{Lm}$ are

\begin{align}
\partial_{t}\hat{\Phi}_{\sigma}^{\dagger}(x,t) & =\sigma v\partial_{x}\hat{\Phi}_{\sigma}^{\dagger}(x,t)
\end{align}
whose general solution in terms of the initial conditions is 
\begin{align}
\hat{\Phi}_{\sigma}^{\dagger}(x,t) & =\hat{\Phi}_{\sigma}^{\dagger}(x+\sigma vt,0)
\end{align}
We see that the soliton operators evolve purely ballistically. This feature
is expected to be valid in any other system in which there exist local
field operators whose evolution satisfies the wave equation, like
for example the massless Thirring model.

Note the distinction we use between the terms ``soliton'' and ``solitary
wave'': the former is used to describe static field configurations
interpolating between different asymptotic vacua, while the latter
describe dynamical fields, i.e. operatorial solutions of the Heisenberg
equation of motion, that correspond to ballistically moving localised
fields. Static solitons may be stable or unstable in which case they
decompose to other excitations, depending on the Hamiltonian that
governs their time evolution. Solitary waves may have no connection
with static solitons or may correspond to clusters of solitons. In
the present case, where the conservation of boson particle number
constrains us to the Hilbert space sector with zero overall solitons,
such excitations appear in the initial state only in pairs of soliton
creation and annihilation operators.

From the above relation we can derive the large time limit of any
fermionic correlation function in terms of initial ones, assuming
that the initial state satisfies translation invariance and cluster
property in terms of the fermions 
\begin{equation}
\lim_{t\to\infty}\llangle[\bigg]\prod_{i=1}^{n}\hat{\Phi}_{\sigma_{i}}^{\dagger}(x_{i},t)\hat{\Phi}_{\sigma_{i}}(y_{i},t)\rrangle[\bigg]=\sum_{\sigma=\pm}\llangle[\bigg]\prod_{i=1}^{n}\hat{\Phi}_{\sigma}^{\dagger}(x_{i})\hat{\Phi}_{\sigma}(y_{i})\rrangle[\bigg]
\end{equation}

Let us now consider the general case of an interacting evolution equation,
like the quantum KdV. The time evolution of the field in the Heisenberg
picture is $\hat{\phi}(x,t)=\mathrm{e}^{\mathrm{i}\hat{H}t}\hat{\phi}(x)e^{-i\hat{H}t}$
where $\hat{\phi}(x)=\hat{\phi}(x,0)$ is the initial field configuration.
Using the general Baker-Campbell-Hausdorff formula, we can formally
express $\hat{\phi}(x,t)$ as an expansion in a basis of initial,
not necessarily local, fields 
\begin{equation}
\hat{\phi}(x,t)=\sum_{\alpha}\int_{-\infty}^{+\infty}dx'\,G_{\alpha}(x-x';t)\hat{\Phi}_{\alpha}(x')
\end{equation}
This formal expansion represents the propagation of field fluctuations
from the initial state up to time $t$. The basis $\{\hat{\Phi}_{\alpha}\}$
is in general infinite and includes derivatives of $\phi$ of any
order, as well as composite operators. Clearly the fields $\hat{\Phi}_{\alpha}$
are not limited to local fields. Even though for a continuous Hamiltonian
involving only contact interactions, the Baker-Campbell-Hausdorff
formula would produce commutators that are seemingly local (derivatives
of the Dirac $\delta$-function of arbitrary high order), such fields
are ill-defined and require normal-ordering or point-splitting that
essentially means that they are non-local. This can be clearly seen
in lattice systems where the general term corresponds to a composite
operator of arbitrarily large length of spatial support. In the case
of a Hamiltonian quadratic in $\hat{\phi}$ and its canonically conjugate
field $\hat{\pi}=\partial_{t}\hat{\phi}$ satisfying the canonical
commutation relation $[\hat{\phi}(x),\hat{\pi}(x')]=i\delta(x-x')$,
the algebra of these operators closes and we obtain a linear evolution
\begin{equation}
\hat{\phi}(x,t)=\int_{-\infty}^{+\infty}dx'\,\left(\dot{G}(x-x';t)\hat{\phi}(x')+G(x-x';t)\hat{\pi}(x')\right)
\end{equation}

Dispersive modes would be associated with time decaying propagators 
$G_{\alpha}(x;t)$. Instead solitary wave solutions of
the Heisenberg equations of motion are defined as localised operator
solutions that evolve ballistically i.e. of the form 
\begin{equation}
\hat{\phi}(x,t)=\hat{\Phi}_{s}(x-v_{s}t-x_{0})
\end{equation}
where $x_{0}$ is a free parameter, the initial position of the solitary wave,
and $v_{s}$ is its velocity. Equivalently, such localised
fields satisfy the dynamical condition 
\begin{equation}
\mathrm{e}^{\mathrm{i}Ht}\hat{\Phi}_{s}(x)\mathrm{e}^{-\mathrm{i}Ht}=\hat{\Phi}_{s}(x-v_{s}t)=\mathrm{e}^{-\mathrm{i}Pv_{s}t}\hat{\Phi}_{s}(x)\mathrm{e}^{\mathrm{i}Pv_{s}t}
\end{equation}
where $\hat{P}$ is the momentum operator, 
generator of spatial translations. The above equation suggests 
that the operators $\hat{\Phi}_{s}(x),\hat{H}$ and
$\hat{P}$ form a closed subalgebra. The localisation condition can
be expressed as exponential decay with the distance of the norm 
of the commutator between the solitary wave operator and the fundamental 
local field
\begin{equation}
\left\Vert \left[\hat{\Phi}_{s}(x),\hat{\phi}(y)\right]\right\Vert <\text{const.}\,\mathrm{e}^{-\lambda_{s}|x-y|}
\end{equation}
We will call such solutions ``quantum solitary waves''. The existence
of such solutions depends on whether quantum corrections preserve
the localisation features of the classical solitary waves or not,
and in general it cannot be excluded from first principles.

Motivated by the semiclassical results for the KdV equation, let us
now consider an analogous large time asymptotic expansion in the quantum
case, i.e. that any initial field configuration $\hat{\phi}_{0}(x)$
evolves so that it eventually splits into a number of quantum solitary
waves $\hat{\Phi}_{s}$ and a time decaying dispersive component $\hat{R}$. 
In this case we would write a formal expansion 
\begin{equation}
\hat{\phi}(x,t)\sim\sum_{s}\hat{\Phi}_{s}(x-x_{0s}-v_{s}t)+\hat{R}(t),\ \text{for }t\to\infty
\end{equation}
where the residual component $\hat{R}(t)$ has extensive spatial support,
but its projection onto a local basis of field operators decays 
uniformly with time. In analogy to the classical case we
may expect that the above sum consists of a single solitary wave only,
since the initial field configuration $\hat{\phi}_{0}(x)$ is a point-like
field. However we do not need to restrict ourselves with this expectation.

\begin{figure}
\includegraphics[width=0.5\paperwidth]{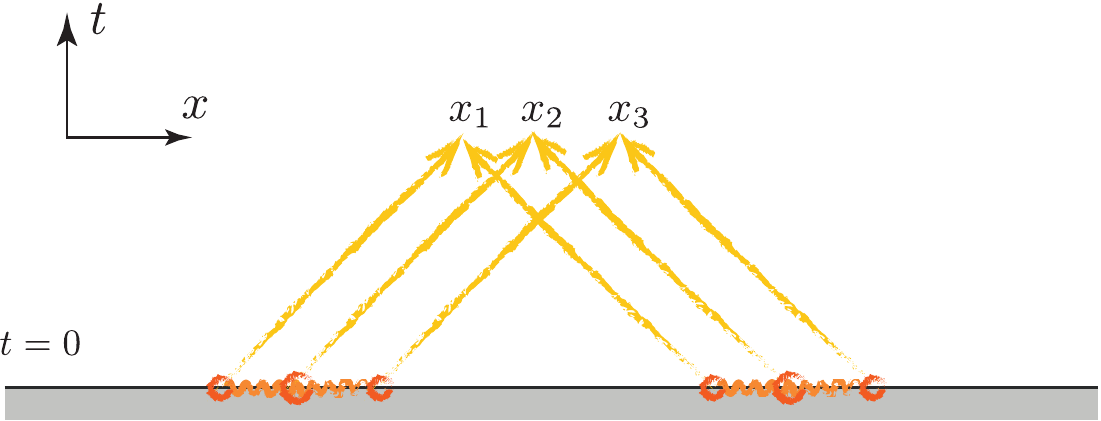}\caption{Equilibration 
in the presence of quantum solitary waves in 1D hydrodynamic systems: Due to their
localised nature and ballistic dynamics, large time correlations between
a set of points are equal to the initial solitary wave field correlations at their
light-cone projections. Due to clustering of initial correlations,
the large time correlations are decomposed into a sum of two components 
originating from the two spatial infinities. \label{fig:concept}}
\end{figure}
By now it should be obvious what the consequences of this scenario
would be for our quench problem. Let us focus on the long time asymptotics
of connected correlation functions of the local field $\hat{\phi}(x,t)$.
Using the above formula and applying it to the time inverted evolution
in order to trace back to the initial fields from which the large
time field $\hat{\phi}(x,t)$ originates, we obtain 
\begin{equation}
\lim_{t\to\infty}\left\langle \prod_{i}\hat{\phi}(x_{i},t)\right\rangle =\lim_{t\to\infty}\left\langle \prod_{i}\sum_{s}\hat{\Phi}_{s}(x_{i}-x_{0s}-v_{s}t)+\hat{R}(t)\right\rangle 
\end{equation}
We can omit the time decaying dispersive part, except for the two
point functions where it does contribute, as in the non-interacting
case. Since the solitary wave operators are spatially localised and moving
away from each other due to their different velocities, from the cluster
property of the initial state the above correlation function splits
into a sum of clusters of same type solitary waves. Lastly using the translational
invariance of the initial state we can remove the coordinate shifts
$-x_{0s}-v_{s}t$. We therefore conclude that
\begin{align}
\lim_{t\to\infty}\left\langle \prod_{i}\hat{\phi}(x_{i},t)\right\rangle  & =\sum_{s}\left\langle \prod_{i}\hat{\Phi}_{s}(x_{i})\right\rangle _{0}
\end{align}
which means that initial correlations of solitary wave operators of all
orders survive up to infinite time (Fig.~\ref{fig:concept}). 
Note that the key features of
the definition of quantum solitary waves that are important in the
above arguments are that they are localised fields, so that their
correlations satisfy the cluster property, and that they are preserved
by time evolution, in particular that the norm of the time-evolved
operators does not reduce with time, as that of the dispersive components.

\section{Discussion of results}\label{sec:discussion}

\subsubsection*{Experimental relevance}

Experimental realisation of quench protocols that are described by
continuous quantum field theories has been achieved in the framework
of 1D split atomic quasi-condensates, which has recently led to the
celebrated first observation of a GGE \cite{exp:GGE}. The theoretical
description of such cold-atom experiments, assuming point-like interactions,
is given by the integrable Lieb-Liniger model. On the other hand,
their low-energy effective description in terms of the atomic wavefunction's
density and phase fields is given by the sine-Gordon model and, in
the gapless phase, the Luttinger model \cite{exp:sG3,exp:sG2}. More specifically
it has been argued that the experimental system can be
prepared in a state that resembles the ground state of the sine-Gordon
model for arbitrary interaction \cite{exp:sine-Gordon}. Basic features
of the general sine-Gordon ground state, like the non-Gaussianity
of correlation functions and the presence of massive excitations and
static solitons, have been observed experimentally. On the other hand, the
evolution Hamiltonian is typically chosen to lie in the gapless regime,
which is described at low-energies by the standard Luttinger model \cite{exp:lightcone1,exp:lightcone2,exp:LL}. In particular
the light-cone form of dynamics as predicted by the Luttinger model
has been observed in \cite{exp:lightcone1}.

Both the sine-Gordon model and the Luttinger model are only low-energy
effective descriptions of the actual system, which may not be expected
to provide exact descriptions of quantum quenches, due to the fact
that instantaneous quenches typically create excitations of arbitrarily
high-energies. One way to suppress the creation of high-energy excitations
so as to remain in the regime of validity of the low-energy approximation,
is to perform the quench over a finite time interval, which
is perhaps a more realistic experimental protocol. In fact the observation
of a light-cone effect, despite the non-relativistic nature of the
system, may be a sign that the presence of high-energy excitations
in the initial state is suppressed. 

On the contrary, in order to take
into account the presence of high-energy excitations, it is necessary
to go beyond the standard Luttinger model description and include
the effects of non-linear dispersion and interactions on the dynamics,
which are captured by quantum hydrodynamic models, like 
the quantum KdV model studied here. Therefore the quantum analogue
of the problem Zabusky and Kruskal considered 50 years ago \cite{ZK}
can now be studied efficiently in cold atom experiments.

Our analysis shows that in the standard Luttinger model, due to the
purely ballistic nature of dynamics, dephasing is prevented and equilibration
is characterised by preservation of the memory of initial non-Gaussian correlations.
In the presence of dispersion or self-interaction of the bosonised
fields, this memory effect breaks down and the steady state is a Gaussian
GGE in terms of the bosonisation or fermionisation fields respectively.
Dispersion of the bosonised fields is a consequence of a non-local
interaction between the original bosons or next-to-leading order contributions
of the kinetic energy in the hydrodynamic equations. On the other
hand, interaction between the bosonised fields corresponds to non-linear
dispersion of the original boson fields. The combination of dispersion and 
interaction gives rise to a new possibility, the formation of solitary waves, 
the stability of which under the effect of quantum fluctuations remains an open problem. 
In the presence of such ballistic localised excitations
the previous memory preservation effect is restored. In this context, 
more specifically in the Lieb-Liniger description of the
experimental system, solitons have been proposed to correspond to
Lieb's second type of excitations which describe hole excitations.
In \cite{exp:GGE} instead a Bogoliubov dispersion
relation is used for the description of the post-quench excitations. 
This type of dispersion is associated with Lieb's first type of fundamental 
excitations that describe phononic
excitations. Assuming that excitations of the second type are suppressed
in the initial state, our analysis suggests that, as a result of the dispersion of 
the phonons, the steady state is Gaussian in terms of the density and phase 
fields.

It is worth mentioning that, in the same way that integrability, as an approximation of real physical
systems, manifests itself on their dynamics as pre-thermalisation
described by a GGE-type equilibration before thermalisation eventually 
takes place, quantum hydrodynamics may also be expected to 
describe the early stages of quench dynamics. In particular, 
considered as an approximation of other integrable models,
quantum hydrodynamical equilibration would correspond to a ``pre-equilibration''
stage, before equilibration to some version of the GGE takes place. 
This scenario relies however on the assumption of a clear separation between the power-law
exponents characterising the temporal decay of corrections at each of these
stages. In fact we have already seen that the low-momentum scaling of the 
dispersion relation of different types of excitations determines the large 
time decay of transients before equilibration is finally reached.

\subsubsection*{Comparison with the GGE and Quench Action method}

Let us now clarify the relation of our arguments and approach with two main 
concepts in the field of out-of-equilibrium quantum physics in 
general integrable models: the GGE and the Quench Action method. 
We have argued that the presence of ballistically moving localised
excitations, like stable quantum solitons, is associated with preservation
of the memory of all their initial correlations at all times. In such
systems, like in the standard Luttinger model discussed here, mere
knowledge of the initial density of excitations is not sufficient
for the complete description of the long time steady state. Nevertheless,
even though solitary waves are one of the principal features of classical
integrability, the existence of such excitations in quantum integrable
systems remains largely unexplored. Adopting the above point of view,
we are led to the conclusion that in principle integrable dynamics
do not necessarily identify with the memory loss of conventional equilibration
and the economy that characterises typical statistical ensembles.
Instead, the decisive factor regarding the information content of
the steady state appears to be the locality properties of the excitations,
which is only indirectly constrained by integrability.

While the GGE has been implicitly associated with extensive memory
loss, since the number of local charges in a typical integrable model
increases only linearly with the system size, in fact its precise
definition is less restricting than that: it is always possible that
there exist additional local or quasi-local charges whose number increases
faster than linearly with the system size and which would capture
all necessary information. In a non-interacting system, like the standard
Luttinger model, such charges would have to be non-quadratic in the
fields so that the large time steady state is non-Gaussian in terms
of them. In this scenario the GGE would be still valid, even though
it would be qualitatively different from its conventional form and
no longer economic: the number of required charges would not increase
simply linearly with the system size as typically in integrable models
but rather exponentially, like the number of clusters between different
points, a feature characteristic of super-integrability or non-Abelian
integrability \cite{non-abelian}, rather than mere integrability.
Moreover, the question of existence of unknown extra charges perhaps
makes this problem of academic rather than practical interest.

In general, although from a fundamental perspective it is important
to understand if the set of local charges is indeed sufficient for
the complete description of the steady state in a general quantum
quench problem, from a practical point of view it is more important
to develop methods to extract partial information about the steady
state (a particular physical observable) from partial information
about the initial state (a set of initial observables). This is even
more essential in systems where the local charges are not explicitly
known or, even if they are, they are ill-defined, as is typically
the case in continuous systems. On the other hand, while explicit
knowledge of the expansion of the initial state in terms of post-quench
eigenstates provides all information needed for the calculation of,
in principle, any local observable, in practice the derivation of such
expansions can be achieved inevitably only for special types of initial
states, like those involving only pairs of oppositely moving excitations
\cite{BEC1,BEC2,KCC,Neel,atr_LL}. The method advocated here circumvents
completely both problems of finding the expansion of the initial state
and of determining the set of suitable charges of the GGE and deriving
its predictions. Instead, our approach is to track the origin of the
information that is required for the derivation of large time observables
from initial ones, by focusing on the long time asymptotics of solutions
of the Heisenberg equations of motion.

Another analytical method for the solution of quantum quench problems in
Bethe Ansatz solvable models, which is largely inspired by statistical
physics concepts, is the so-called Quench Action or representative
state method \cite{QA,QA2}. In this approach the initial state is
described in the thermodynamic limit by a representative state and
nearby excitations \cite{DNPC}. Under the assumption of dephasing
during the time evolution, the effect of such excitations decays with
time, so that in the large time limit local observables are expected
to be completely described by the representative state only. According
to this approach the large time steady state for a non-interacting
model is predicted to be Gaussian and only information about the density
of momentum excitations is expected to survive at large times, as
long as we restrict ourselves to local observables. We have demonstrated
that this is not true for systems with non-interacting dynamics characterised
by linear dispersion, where essentially all information about non-Gaussian
initial correlations is transferred intact up to infinite times. The
reason is quite simple: one of the conditions of the quench action
method is that the evolution induces dephasing that wipes out all
information about separate clusters of initial excitations, which
is obviously not valid here. More specifically, even if it is true
that the initial state can be completely described by a representative
state and the set of adjacent excitations, in the absence of dephasing
all such excitations remain present in the steady state and therefore
the representative state alone does not provide a complete description. %

In general, the physical conditions on the dynamics that would guarantee
sufficient dephasing in order for the quench action method to be applicable
remain to be specified. While the linear dispersion considered above
is an idealised and rather unrealistic case, as we have argued the
same memory effect is expected to be present in systems exhibiting
stable solitons, a genuine feature of classical integrability albeit elusive
in the framework of the Bethe Ansatz solution for quantum integrable models.
Therefore it remains an interesting open question to test if sufficient dephasing
takes place in general, especially when the initial state does not simply consist of opposite
momentum pair excitations \cite{BEC1,BEC2,KCC,Neel,atr_LL}. Such
states are special and analogous to Gaussian states in non-interacting
systems, in the sense that they are fully determined by their density
of momentum excitations, for which reason they are not so suitable
for the study of memory loss.

\section{Conclusions \& Outlook}\label{sec:open_questions}

We have studied quench dynamics and equilibration in quantum hydrodynamic
systems. Equilibration in such systems presents special interest
since, due to the gaplessness of the evolution Hamiltonian, the stationary
phase argument which is the typical explanation of dephasing in quantum
systems is not directly applicable. Moreover, as we have shown, the
low-energy description of such systems as given by the standard Luttinger
model exhibits purely ballistic dynamics which prevents dephasing
so that equilibration preserves the memory of non-Gaussian initial correlations. 
However, while the gaplessness of the energy spectrum appears
to be a necessary condition for this memory effect, it is not a sufficient
one. What we have shown in the present work is that this effect
is actually a consequence of a very delicate property of the Luttinger model,
its linear dispersion relation. In fact, the presence of either dispersion
or chiral interactions, which are irrelevant at equilibrium, 
breaks down this effect and induces dephasing
so that the steady state is given by a Gaussian GGE in terms of the
bosonisation or fermionisation fields respectively. In the case of
simultaneous presence of both dispersion and chiral interactions,
the interplay between these two effects and their role on equilibration
remains an open problem, since the potential formation of dynamically 
stable quantum solitons would restore the ballistic localised character of dynamics
and the memory preservation effect. Following the connection 
between the nature of equilibration and transport properties introduced 
in \cite{Eisert16}, we have identified quantum solitons as carriers 
of information of initial correlations in quantum quenches, opening
a new perspective to the ongoing research on the stability 
of soliton excitations in 1D quantum systems 
\cite{q_hydrodyn1,QH0,QH1,QH2,QH3,QH4}.

{Although we have focused on aspects of quench dynamics and equilibration
in 1D quantum hydrodynamics, our analysis has consequences
on several physical problems in the broader context of quantum out-of-equilibrium
physics. 
First of all, our analysis of the effects of dispersion provides a number of conclusions
about the application of \emph{Luttinger liquid theory} to quantum quenches
\cite{Cazalilla,s-jstat3}. It is clear that the standard Luttinger
model does not lead to Gaussian relaxation in terms of the bosonised
fields, since a non-Gaussian initial state would remain such for all
times due to the linear dispersion relation. (It is still possible
however that an out-of-equilibrium version of the ``Luttinger liquid
conjecture'' for the relation between exponents of correlation functions, analogous
to that at equilibrium \cite{LL_conj}, may still hold out of equilibrium.) 
In order to study quench dynamics and equilibration it is 
therefore necessary to go beyond the standard
Luttinger model with linear dispersion and take into account e.g.
dispersion effects which are the result of a non-local interaction
or next-to-leading order contributions of the kinetic energy \cite{Cazalilla}.
In this case, the steady state correlations depend significantly on the 
nonlinearity of the dispersion relation, as shown in \eqref{disp_2pt-cf}, even though the latter
does not affect their large distance asymptotics. These remarks may
explain the minor discrepancies between numerics and Luttinger liquid
predictions in spin chains observed in \cite{CCE}.}

{While we focused our analysis on the 
quantum KdV equation, which provides a general low-energy effective description of 
the one-dimensional interacting Bose gas, the latter can be faithfully described
also by the \emph{Lieb-Liniger model}, an integrable model of non-relativistic
bosons with contact interactions that can be solved exactly by means
of the Bethe Ansatz. 
In the case of repulsive interaction, evolution under the Lieb-Liniger
Hamiltonian is expected to lead to dispersion of the KdV solitons,
at least on the basis of semiclassical calculations \cite{qKdV-LL}.
It is worth mentioning that the soliton dispersion relation is different
from that of phonons and would lead to a slower dispersion as $t^{-5/2}$.
It can be therefore expected that it is exactly the soliton excitations
that govern the large time scaling of the corrections to equilibration. 
In the attractive case, on the other hand, such localised
solutions are expected to be more stable and may be linked to many
particle bound states. A challenging question is to study 
the soliton stability in the attractive Lieb-Liniger model and
their effect on equilibration after a quench starting from a general 
initial state.}

{Our analysis of the dispersionless interacting KdV applies equally well to 
general quantum quenches to the \emph{Tonks-Girardeau 
limit} of quantum gases, since in this case the dynamics is described 
essentially by the same free fermion Hamiltonian. We have shown 
that, at least under the assumption of clustering of fermionic correlations
in the initial state, the large time density correlations are Gaussian, 
thus generalising results of \cite{KCC} for the BEC state. 
However, as we explained, the condition of initial clustering of fermionic 
correlations is non-trivial, due to the non-local nature of the Jordan-Wigner 
transformation. We have shown that clustering is valid for composite 
fermionic operators of even number and we have given some arguments
that the exponential clustering observed for the BEC state \cite{KCC}
is expected to be more generally valid. 
}

{Evidence for the existence of dynamically stable quantum solitons
comes mainly from continuous rather than \emph{lattice systems}. Indeed,
in the most prominent examples of integrable spin chains, like the
XXZ model, dispersion seems to always dominate, so that initially
prepared static solitons spread and appear to decay with time when
they are let to evolve \cite{spin-chain_solitons}. It is an interesting
question if this type of behaviour holds generally as a consequence
of some more fundamental property. For example, a general feature
of the dynamics of spin chains with short (finite) range interactions
is the \emph{Lieb-Robinson bound} \cite{LR,Lieb-Robinson,LR_BHV}, i.e. the
existence of a maximum velocity of signal propagation. This bound
constrains the dynamics and the excitation spectrum, so that it may
obstruct the emergence of stable solitons. Long-range lattice systems
instead, similarly to continuous non-relativistic systems, do not exhibit
such constraints on their dynamics.}

{Our analysis may also be relevant to the problem of 
\emph{quantum transport} in integrable models, as in the case of 
quenches starting from inhomogeneous initial states.
In this context, a (classical) generalised hydrodynamic behaviour has been recently 
shown to emerge \cite{hydrodynamics1,hydrodynamics2}. Quantum hydrodynamics 
is conceptually and in principle different from the latter approach, which refers to 
hydrodynamic equations satisfied by, not the quantum fields themselves, 
but rather their expectation values. 
An open question is whether the emergence of generalised hydrodynamics implies 
the validity of underlying quantum hydrodynamics as well.
In this case, the presence of stable quantum solitons is expected to affect 
the steady state values of higher order cumulants of the currents. 
}

{On the more mathematical side, an important open question 
is that of the existence and stability of solitary waves in quantum systems 
which can be studied by going beyond the semiclassical approximation. 
Some arguments analogous to the 
semiclassical analysis come from the \emph{collective field theory} 
of Jevicki and Sakita \cite{JS,Jevicki} as applied to
the Calogero model, which is described by the Benjamin-Ono hydrodynamic
equation. 
On the other hand, an exact treatment could be based on the application
of the \emph{Quantum Inverse Scattering Method} \cite{Faddeev,thacker},
especially as a means to derive the general solution of the Heisenberg
equations of motion in infinite space through the \emph{Gelfand-Levitan-Marchenko}
equations and the \emph{Rosales series} solution \cite{IST_quench-map}.
To the best of our knowledge this approach has not been applied to the
quantum KdV equation 
\footnote{Remarkably, while the classical KdV equation was historically the first to be
recognised as solvable by means of the inverse scattering method \cite{KdV_solution,ZF_QISM_KdV},
the development of the quantum version of this method followed a different
path and was applied first to the sine-Gordon model and the nonlinear
Schr\"{o}dinger equation, overlooking the quantum KdV equation. As explained
in \cite{history}, this was so because ``the KdV equation look{[}ed{]}
irrelevant from the quantum point of view''.}. %
Another approach may be based on the interpretation of the quantum
KdV equation as a \emph{Riemann-Hilbert problem} for an operator-valued field,
i.e. the problem of determining an analytic field inside a domain
when matching conditions are given along its boundary. Such an approach
has been applied to the \emph{quantum Benjamin-Ono equation} in \cite{q_hydrodyn1}.
In analogy to the classical case, the long time asymptotics may then
be obtained using a quantum version of the \emph{non-linear steepest descent
method} \cite{math:KdV_asympt}.}

\begin{acknowledgments}
I am grateful to F. Franchini, M. A. Cazalilla and B. Doyon for inspiring discussions. I acknowledge
support from the A{*}MIDEX project Hypathie (no. ANR-11-IDEX- 0001-02)
funded by the ``Investissements d'Avenir'' French
Government program, managed by the French National Research Agency
(ANR).\end{acknowledgments}




\appendix

\section{Derivation of scaling relations in the dispersive case}\label{app:dispersion}

In this appendix we present a detailed calculation of connected correlation functions
of field derivatives in the presence of dispersion of the bosonisation fields.
We show that, as a result of dispersion,
they vanish at large times for all orders larger than two, i.e. the
large time limit is described by a Gaussian state in terms of the
bosonisation fields. 

Let us start with the two-point function, which
we express in terms of the initial one 
\begin{align}
\llangle[\bigg]\partial_{x}\hat{\vartheta}(x,t)\partial_{y}\hat{\vartheta}(y,t)\rrangle[\bigg] & =\int dx'dy'\,\sum_{\nu,\mu=0,1}G^{(\nu,1-\nu)}(x-x',t)G^{(\mu,1-\mu)}(y-y',t)C_{0}^{(\nu,\mu)}(x'-y')\nonumber\\
 & =\int\frac{dk}{2\pi}\,\mathrm{e}^{\mathrm{i}k(x-y)}\sum_{\nu,\mu=0,1}\tilde{G}^{(\nu,1-\nu)}(k,t)\tilde{G}^{(\mu,1-\mu)}(-k,t)\tilde{C}_{0}^{(\nu,\mu)}(k)
\end{align}
where 
\begin{equation}
C_{0}^{(\nu,\mu)}(x-y)=\llangle[\bigg]\hat{\vartheta}^{(1-\nu,\nu)}(x)\hat{\vartheta}^{(1-\mu,\mu)}(y)\rrangle[\bigg]
\end{equation}
are the initial two-point correlation functions of field derivatives
and in the Fourier transform we have explicitly factorised the momentum
conserving $\delta$-function. In the large time limit, the oscillating
terms that result from the expressions \eqref{dtG} and \eqref{dxG} 
decay with time, provided that the initial two-point correlations
exhibit no divergences for small momenta, which is ensured by sufficiently
strong initial clustering (exponential or sufficiently strong algebraic
decay with the distance). We therefore obtain 
\begin{align}
\lim_{t\to\infty}\llangle[\bigg]\partial_{x}\hat{\vartheta}(x,t)\partial_{y}\hat{\vartheta}(y,t)\rrangle[\bigg] & =\frac{1}{2}\int\frac{dk}{2\pi}\,\mathrm{e}^{\mathrm{i}k(x-y)}\left(\tilde{C}_{0}^{(0,0)}(k)+\frac{1}{v^{2}f^{2}(k)}\tilde{C}_{0}^{(1,1)}(k)\right) \label{disp_2pt-cf_a}
\end{align}

For higher order connected correlation functions $n>2$, we have
\begin{align}
\llangle[\bigg]\prod_{i=1}^{n}\partial_{x}\hat{\vartheta}(x_{i},t)\rrangle[\bigg] & =\int\prod_{i=1}^{n}dx_{i}'\,\prod_{i=1}^{n}\sum_{\nu_{i}=0,1}G^{(\nu_{i},1-\nu_{i})}(x_{i}-x'_{i},t)C_{0}^{\{\nu_{i}\}}(\{x'_{i}\})\nonumber\\
 & = \int\prod_{i=1}^{n}dx'_{i}\,\prod_{i=1}^{n}\sum_{\sigma_{i}=\pm}\sum_{\nu_{i}=0,1}\sigma_i^{\nu_{i}}\Delta^{(\nu_{i})}(x'_{i};\sigma_i t)C_{0}^{\{\nu_{i}\}}(\{x_{i}-x'_{i}+\sigma_i vt\})\label{eq:corr-time-evol}
\end{align}
where 
\begin{equation}
C_{0}^{\{\nu_{i}\}}(\{x_{i}\})=\llangle[\bigg]\prod_{i=1}^{n}\hat{\vartheta}^{(1-\nu_{i},\nu_{i})}(x{}_{i})\rrangle[\bigg]
\end{equation}
are initial connected correlation functions between local fields. 

The analysis of the long time asymptotics can be done easier 
in Fourier space, using standard results from the theory of 
multiple Fourier transforms \cite{book:asymptotics}. In Fourier space, 
\eqref{eq:corr-time-evol} can be written explicitly as
\begin{align}
& \llangle[\bigg]\prod_{i=1}^{n}\partial_{x}\hat{\vartheta}(x_{i},t)\rrangle[\bigg] =\int\prod_{i=1}^{n}\frac{dk_{i}}{2\pi}\,\mathrm{e}^{\mathrm{i}\sum_{i}k_{i}x_{i}}\prod_{i=1}^{n}\sum_{\nu_{i}=0,1}\tilde{G}^{(\nu_{i},1-\nu_{i})}(k_{i},t)\delta({\textstyle \sum_{i}k_{i}})\tilde{C}_{0}^{\{\nu_{i}\}}(\{k_{i}\})\nonumber\\
 & \quad =\frac{1}{2^{n}}\int\prod_{i=1}^{n}\frac{dk_{i}}{2\pi}\,\mathrm{e}^{\mathrm{i}\sum_{i}k_{i}x_{i}}\sum_{\sigma_{i}=\pm}\sum_{\nu_{i}=0,1}\sigma_{i}^{\nu_{i}}\mathrm{e}^{\mathrm{i}vt\sum_{i}\sigma_{i}k_{i}}\mathrm{e}^{\mathrm{i}vt\sum_{i}\sigma_{i}k_{i}(f(k_{i})-1)}\prod_{i=1}^{n}\frac{1}{\left(vf(k_{i})\right)^{\nu_{i}}}\delta({\textstyle \sum_{i}k_{i}})\tilde{C}_{0}^{\{\nu_{i}\}}(\{k_{i}\})
\end{align}
According to the cluster decomposition property, the Fourier transform
of the initial connected correlation functions 
contains a momentum conservation $\delta$-function singularity, which
we have explicitly factorised, but no other such singularity. 
We now observe that, first of all, the leading large time corrections come from
the two contributions of initial correlations at the left and right
spatial infinities. These correspond to the two terms of the above
sum for which all $\sigma_{i}$ are equal to either 1 or -1, i.e.
\begin{align}
\lim_{t\to\infty}\llangle[\bigg]\prod_{i=1}^{n}\partial_{x}\hat{\vartheta}(x_{i},t)\rrangle[\bigg] & \sim\frac{1}{2^{n}}\sum_{\sigma=\pm}\int\prod_{i=1}^{n}\frac{dk_{i}}{2\pi}\,\mathrm{e}^{\mathrm{i}\sum_{i}k_{i}x_{i}}\sum_{\nu_{i}=0,1}\sigma^{\nu_{i}}\mathrm{e}^{\mathrm{i}\sigma vt\sum_{i}k_{i}(f(k_{i})-1)}\prod_{i=1}^{n}\frac{1}{\left(vf(k_{i})\right)^{\nu_{i}}}\delta({\textstyle \sum_{i}k_{i}})\tilde{C}_{0}^{\{\nu_{i}\}}(\{k_{i}\})
\end{align}
For these terms the factor $\mathrm{e}^{\mathrm{i}vt\sum_{i}\sigma_{i}k_{i}}$
disappears due to the $\delta({\textstyle \sum_{i}k_{i}})$ and the asymptotics is 
determined by the dispersive scaling
law $\mathrm{e}^{\mathrm{i}\sigma vt\sum_{i}k_{i}(f(k_{i})-1)}$. On the contrary, for
all other terms additional multiplicative factors $\sim1/t$ enter
in the asymptotics and therefore they decay much faster with time.
Intuitively the reason for this decomposition into two contributions
from the two spatial infinities is that the support of the propagator
is centered at the two light-cone projection points which move away
from each other faster than its spreading.

From the above formula, by rescaling the integration variables as
$k_{i}\to t^{-1/\gamma}k_{i}$ and assuming exponential initial clustering,
in which case connected correlation functions have no singularity
for real small momenta, we obtain a general scaling law 
\begin{equation}
\lim_{t\to\infty}\llangle[\bigg]\prod_{i=1}^{n}\partial_{x}\hat{\vartheta}(x_{i},t)\rrangle[\bigg]\sim\frac{1}{t^{(n-1)/\gamma}}F(\{x_{i}/t^{1/\gamma}\})
\label{scaling1_a}
\end{equation}
where $F(\{s_{i}\})$ has a finite non-zero value when all of its arguments
are zero. For algebraic initial clustering, additional multiplicative
factors are expected, depending on the low-momentum behaviour of initial
correlations.

\section{Cluster decomposition of fermionic correlations}\label{app:fermionic_clustering}

\begin{figure}
\includegraphics[width=0.4\paperwidth]{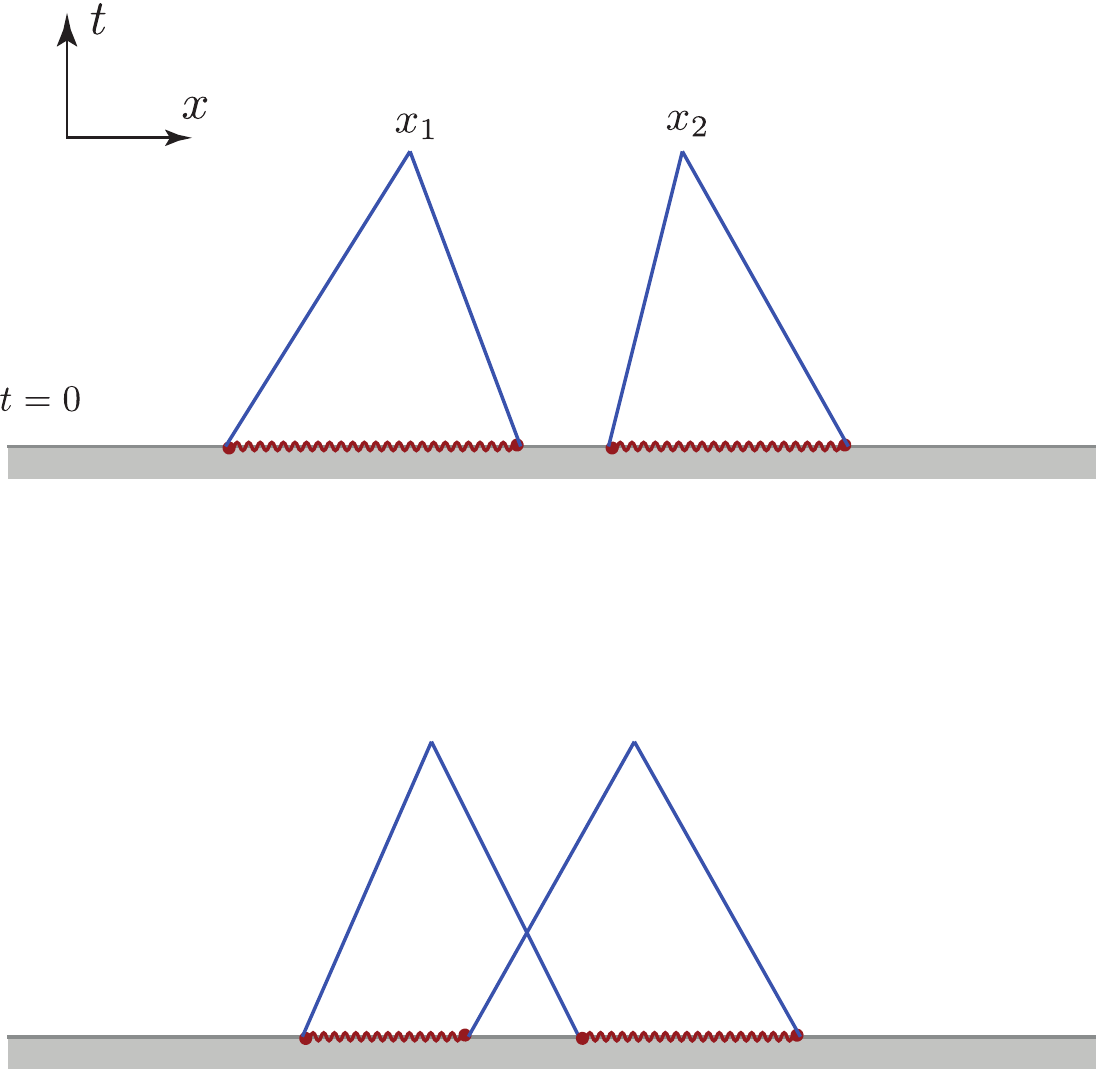}
\caption{The effect of interaction can be seen in the fermionised
representation as dispersion. 
Large time correlations are determined by the large distance asymptotics
of Jordan-Wigner string correlations, which despite their non-local nature,
satisfy the clustering property due to the cancellation of overlapping strings.
\label{fig:strings}}
\end{figure}

In this appendix we analyse the condition of cluster decomposition property of 
fermionic correlation functions in a state that exhibits this property for bosonic correlation functions. Using the transformation \eqref{JW} relating the original boson and fermion fields, which is essentially a Jordan-Wigner transformation, we can write pairs of fermion operators as string operators
\begin{align}
\hat{\Phi}_\sigma^{\dagger}(s)\hat{\Phi}_\sigma(s') & \sim
\exp\left(\mathrm{i}\pi\int_{\min\{s,s'\}}^{\max\{s,s'\}}\mathrm{d}y\,\hat{\rho}_\sigma(y)\right)\nonumber\\
& =\,\,:\exp\left(-2\int_{\min\{s,s'\}}^{\max\{s,s'\}}\mathrm{d}y\,\hat{\rho}_\sigma(y)\right):
\end{align}
where in the last form we used the normal ordering formula $\mathrm{e}^{\lambda A}=\;:\mathrel\mathrm{e}^{(\mathrm{e}^{\lambda}-1)A}:$
\cite{Grosse}. From the latter relation, it can be easily seen that, 
despite the fact that the string operators are non-local, the
following clustering property holds for any translationally invariant 
state that is local in terms of the chiral field derivatives, that is, 
in terms of the original boson field
\begin{equation}
\lim_{R\to\infty}\left\langle \prod_{i=1}^{2n}\hat{\Phi}^{s_{i}}(x_{i})\prod_{j=1}^{2m}\hat{\Phi}^{s_{j}}(x{}_{j}+R)\right\rangle =\left\langle \prod_{i=1}^{2n}\hat{\Phi}^{s_{i}}(x_{i})\right\rangle \left\langle \prod_{j=1}^{2m}\hat{\Phi}^{s_{j}}(x{}_{j})\right\rangle 
\end{equation}
The latter property means that correlations of fermionic operators factorise when
the points are split into two subsets of even number. 
(In the above the superscripts $s_i$ denote creation or annihilation operators). 
This is because the Jordan-Wigner 
strings always connect operators that are successive in spatial ordering,
since, when two strings overlap in space, they cancel in their common
overlap (Fig. \ref{fig:strings}). Therefore separating the set of coordinates 
in subsets of even number always results in two extended operators that 
are however supported in finite spatial intervals. 
In fact the above property has 
been explicitly used in the special case of a BEC initial state considered
in \cite{KCC}. 

The non-trivial case is when the set of points is split into subsets
of odd number, since then there is one infinitely long string connecting
the two subsets (the rightmost operator of the left subset to the
leftmost operator of the right subset). While it may be true that
clustering of correlations of the local boson field $\hat{\Psi}$
implies also clustering of correlations of the non-local fermion field
$\hat{\Phi}$, we will not address this question here. We should mention
however that in the only case of initial state for which fermionic
correlations have been explicitly calculated, the BEC state, fermionic
correlations turn out to satisfy exponential clustering \cite{KCC}.
This is expected to be more generally valid, as we can see by
writing the expectation value of the string operator as a cluster
expansion 
\begin{align}
\left\langle \exp\left(\mathrm{i}\pi\int_{x}^{y}\mathrm{d}s\,\hat{\rho}(s)\right)\right\rangle  & =\left\langle :\exp\left(-2\int_{x}^{y}\mathrm{d}s\,\hat{\rho}(s)\right):\right\rangle \nonumber\\
 & =\exp\left(-2\rho_{0}|x-y|+\sum_{n=2}^{\infty}\frac{(-2)^{n}}{n!}\prod_{i=1}^{n}\int_{x}^{y}ds_{i}\,C_{0}^{\rho}(s_{1},s_{2},...,s_{n})\right)
\end{align}
where 
\begin{align}
C_{0}^{\rho}(x_{1},x_{2},...,x_{n}) & =\left\langle :\rho(x_{1})\rho(x_{2})...\rho(x_{n}):\right\rangle 
\end{align}
are the normal-ordered connected correlation functions of the density
operator. At large distances we find 
\begin{align}
\left\langle \exp\left(\mathrm{i}\pi\int_{x}^{y}\mathrm{d}s\,\hat{\rho}(s)\right)\right\rangle  & \sim\exp\left[-\left(2\rho_{0}+\sum_{n=2}^{\infty}\frac{(-\text{1})^{n+1}}{n!}\prod_{i=2}^{n}\int_{-\infty}^{+\infty}ds_{i}\,C_{0}^{\rho}(0,s_{2},...,s_{n})\right)|x-y|\right],\quad\text{for }|x-y|\to\infty
\end{align}
The convergence of the integrals in the thermodynamic limit is guaranteed
by a sufficiently strong cluster property. Note that this time the
required cluster property refers to correlations of the density which
is a local field. On ground states of 1D quantum gases the large distance
decay of connected density correlations is algebraic. Following a
standard result of Luttinger liquid theory \cite{Haldane}, the decay
of the density two-point correlation function is proportional to the
inverse square of the distance. Similarly, for higher order correlation
functions Luttinger liquid theory gives at most inverse square decay.
Therefore each of the above integrals is convergent in ground states
of Luttinger liquids. On the other hand, the convergence of the series
can be expected to be true at least for sufficiently low densities,
since $C_{0}^{\rho}(x_{1},x_{2},...,x_{n})\sim\rho_{0}^{n}$ so the
series can be seen as a Taylor expansion in the density.

\section{The classical KdV equation and its general solution}\label{app:classical_KdV}

In this appendix we give an overview of the general solution 
of the classical KdV equation and its long time asymptotics. 
The KdV equation in its standard form is 
\begin{equation}
\partial_{t}u(x,t)=6u(x,t)\partial_{x}u(x,t)-\partial_{x}^{3}u(x,t)
\end{equation}
As is well-known, this integrable differential equation admits solitary wave solutions. 
The one-soliton solution is 
\begin{equation}
u_{s}(x,t)=-\frac{1}{2}\frac{v_{s}}{\cosh^{2}\left[\tfrac{1}{2}\sqrt{v_{s}}\left(x-v_{s}t-x_{0}\right)\right]}
\end{equation}
where $v_{s}>0$ is the soliton velocity and $x_{0}$ is arbitrary.
Multi-soliton solutions also exist, as well as dispersive wave solutions.

The general solution to this equation 
for an arbitrary initial condition is given by the inverse scattering
transform, which was actually discovered and first applied to precisely 
this equation \cite{KdV_solution}. The inverse scattering approach
reduces the above nonlinear equation to two linear problems: finding
the spectrum of the Schr\"{o}dinger equation with an external potential
(direct scattering problem) and inversely finding the potential that
corresponds to a given spectrum (inverse-scattering problem). In more
detail the solution steps are as follows: 
\begin{enumerate}
\item From the initial condition $u(x,0)$ derive the scattering data of
the associated Schrödinger equation 
\begin{equation}
-\frac{d^{2}}{dx^{2}}\psi+u(x,0)\psi=E\psi
\end{equation}
with potential $V(x)=u(x,0)$, i.e. the discrete spectrum $E=-\kappa_{n}^{2}$
corresponding to bound states of the Schr\"{o}dinger equation, the associated
norming (normalisation) constants $c_{n}$ and the reflection coefficient
$R(k)$ of the continuous spectrum. 
\item Time evolve the scattering data: the discrete spectrum eigenvalues
$\kappa_{n}$ remain time invariant, while $c_{n}(t)=c_{n}(0)\mathrm{e}^{A\kappa_{n}^{3}t}$
and $R(k,t)=R(k,0)\mathrm{e}^{\mathrm{i}Bk^{3}t}$. 
\item Solve the inverse scattering problem of deriving the time evolved
solution $u(x,t)$ as the potential of a Schrödinger equation associated
with the time evolved scattering data. 
\end{enumerate}%
\begin{figure}
\includegraphics[width=0.7\paperwidth]{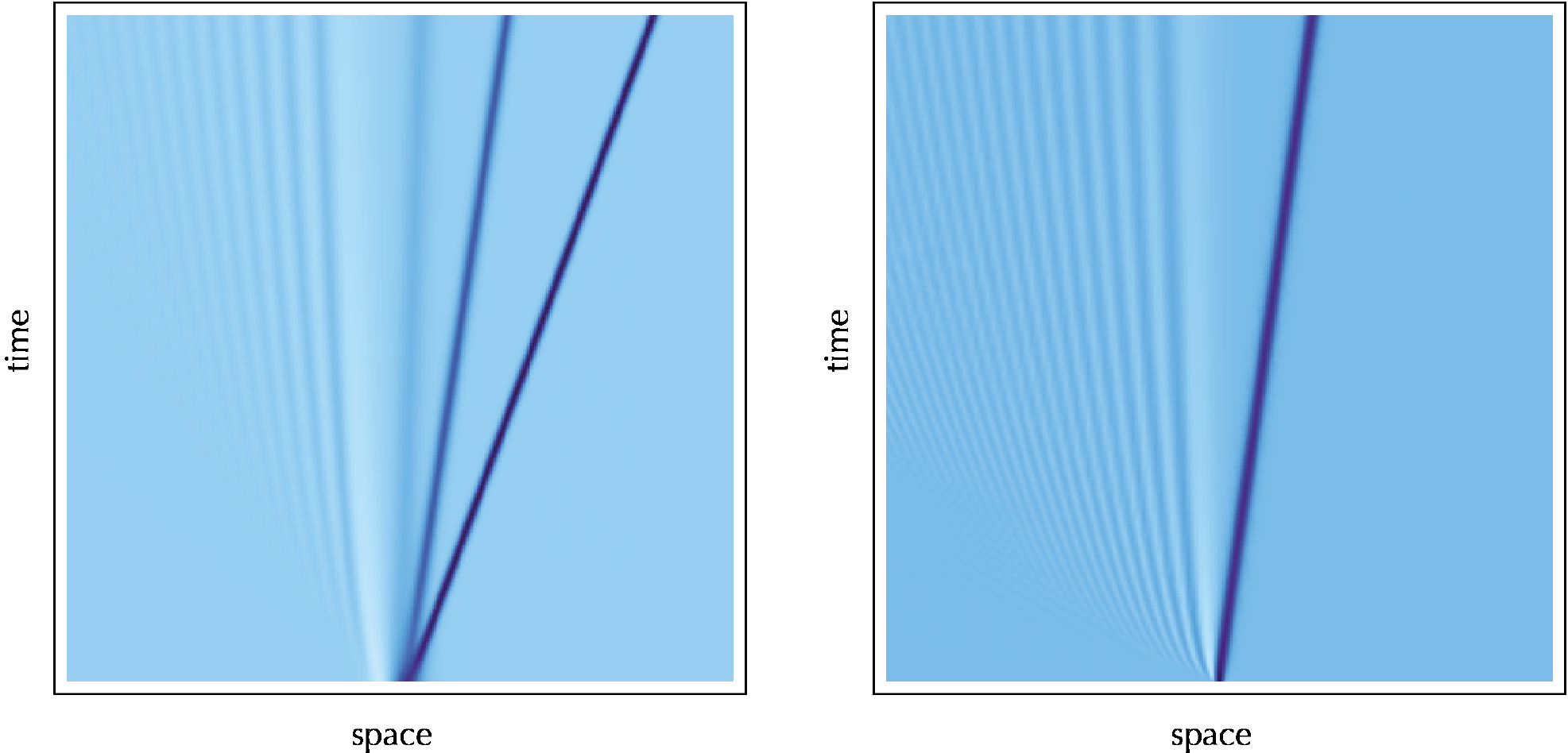}\caption{Time evolution 
of an initial fluctuation following the classical
KdV equation. Two different initial conditions are considered: In
the general case (left) the evolution leads to decomposition into
a number of solitons travelling to one direction and dispersive modes
travelling to the opposite. For a single narrowly-localised pulse
(right) there is precisely one soliton.\label{fig:KdV_sol}}
\end{figure} %
Discrete eigenvalues of the Schr\"{o}dinger equation correspond to solitary waves, 
while the continuum part of the spectrum corresponds to dispersive modes. 
Note that there is at least one discrete eigenvalue for any negative
initial condition $u(x,0)$. In the limit of an infinitely narrow
localised initial configuration, the Dirac $\delta$-pulse $u(x,0)=-\lambda\delta(x)$,
there is precisely one bound state eigenvalue $\kappa=\frac{1}{2}\lambda$.

What we are interested in however is not a formal expression for the
general solution, but rather the general form of the long time asymptotics
for any initial condition. The power of integrability can be seen
in the fact that this partial but most essential information about
the solution can be derived without explicitly knowing it. Indeed
using the nonlinear steepest descent method, it has be shown that
at large times the general solution for any initial condition that
is sufficiently fast decaying at spatial infinities, decomposes to
a superposition of a finite number of solitons moving at different
velocities to the same direction and a decaying in time dispersive
part moving to the opposite direction \cite{math:KdV_asympt} (Fig.
\ref{fig:KdV_sol}) 
\begin{equation}
u(x,t)\sim\begin{cases}
-2\sum_{j=1}^{N}\frac{\kappa_{j}^{2}}{\cosh^{2}(\kappa_{j}x-4\kappa_{j}^{3}t-a_{j})} & ,\,x/t>0\\
O\left(t^{-1/2}\right) & ,\,x/t<0
\end{cases}\quad\text{for }t\to\infty,\,x/t\text{ fixed}
\end{equation}
for some suitable phase shifts $a_{j}$.

\end{document}